\documentclass[12pt, draftclsnofoot, onecolumn]{IEEEtran}
\usepackage{amssymb}
\usepackage{amsmath}
\usepackage{cite}
\usepackage{url}
\usepackage{xcolor}
\usepackage{cite,graphicx,amsmath,amssymb}
\usepackage{subfigure}
\usepackage{citesort}
\usepackage{fancyhdr}
\usepackage{mdwmath}
\usepackage{mdwtab}
\usepackage{caption}
\usepackage{amsthm}
\usepackage{setspace}
\usepackage{algorithm}
\usepackage{algorithmic}

\newtheorem{remark}{Remark}
\newtheorem{theorem}{Theorem}

\newtheorem{lemma}{Lemma}

\newtheorem{corollary}{Corollary}

\newtheorem{proposition}{Proposition}

\allowdisplaybreaks
\begin{document}

\title{Exploiting Intelligent Reflecting Surfaces in NOMA Networks: Joint Beamforming Optimization}
%

\author{

Xidong~Mu,~\IEEEmembership{Student Member,~IEEE,}
        Yuanwei~Liu,~\IEEEmembership{Senior Member,~IEEE,}
       Li~Guo,~\IEEEmembership{Member,~IEEE,}
       Jiaru~Lin,~\IEEEmembership{Member,~IEEE,}
       and Naofal~Al-Dhahir,~\IEEEmembership{Fellow,~IEEE}

\thanks{Part of this work has been submitted to the IEEE Global Communications Conference (GLOBECOM), Taipei, Taiwan, Dec 7-11, 2020.~\cite{Mu2020}}
\thanks{X. Mu, L. Guo, and J. Lin are with School of Artificial Intelligence and Key Laboratory of Universal Wireless Communications, Ministry of Education, Beijing University of Posts and Telecommunications, Beijing, China. (email:\{muxidong, guoli, jrlin\}@bupt.edu.cn).}
\thanks{Y. Liu is with the School of Electronic Engineering and Computer Science, Queen Mary University of London, London, UK. (email:yuanwei.liu@qmul.ac.uk).}
\thanks{N. Al-Dhahir is with the Department of Electrical and Computer Engineering, The University of Texas at Dallas, Richardson, TX 75080 USA.(e-mail: aldhahir@utdallas.edu).}
}

\maketitle
\vspace{-1.5cm}
\begin{abstract}
\vspace{-0.5cm}
This paper investigates a downlink multiple-input single-output intelligent reflecting surface (IRS) aided non-orthogonal multiple access (NOMA) system, where a base station (BS) serves multiple users with the aid of IRSs. Our goal is to maximize the sum rate of all users by jointly optimizing the active beamforming at the BS and the passive beamforming at the IRS, subject to successive interference cancellation decoding rate conditions and IRS reflecting elements constraints. In term of the characteristics of reflection amplitudes and phase shifts, we consider ideal and non-ideal IRS assumptions. To tackle the formulated non-convex problems, we propose efficient algorithms by invoking alternating optimization, which design the active beamforming and passive beamforming alternately. For the ideal IRS scenario, the two subproblems are solved by invoking the successive convex approximation technique. For the non-ideal IRS scenario, constant modulus IRS elements are further divided into continuous phase shifts and discrete phase shifts. To tackle the passive beamforming problem with continuous phase shifts, a novel algorithm is developed by utilizing the sequential rank-one constraint relaxation approach, which is guaranteed to find a locally optimal rank-one solution. Then, a quantization-based scheme is proposed for discrete phase shifts. Finally, numerical results illustrate that: i) the system sum rate can be significantly improved by deploying the IRS with the proposed algorithms; ii) 3-bit phase shifters are capable of achieving almost the same performance as the ideal IRS; iii) the proposed IRS-aided NOMA systems achieve higher system sum rate than the IRS-aided orthogonal multiple access system.
\end{abstract}
\section{Introduction}
Recently, intelligent reflecting surface (IRS) aided transmission has drawn significant attention due to its superior performance in enhancing spectrum efficiency (SE) and energy efficiency (EE) for wireless communication networks~\cite{Renzo2019Smart,Basar}. \textcolor{black}{An IRS is a planar array comprising a large number of low cost passive reflecting elements, which can passively reflect the incident electromagnetic wave and simultaneously change its amplitude and phase shift~\cite{Wu_MAG}.} With this new degree-of-freedom (DoF), an IRS can be deployed to create an additional reflection link when the transmitter and receiver are blocked by obstacles. \textcolor{black}{Compared with conventional relaying technologies (e.g amplify-and-forward (AF) and decode-and-forward (DF) relays), IRSs require much less energy consumption due to the nearly passive characteristics~\cite{Huang_IRS_ax}.} \textcolor{black}{Therefore, IRS technology has drawn tremendous attention from both academia and industry, and has been regarded as a promising solution in future 6G networks~\cite{Huang_IRS_ax,Zhao_survey}.}\\
\indent Non-orthogonal multiple access (NOMA) is one of the key technologies in future wireless communication networks due to its higher SE achievement, user fairness guarantee, and massive connectivity support~\cite{Liu2017,Cai_survey}. The key idea of NOMA is to serve multiple users in the same resource block (i.e., time, frequency and code), where superposition coding (SC) and successive interference cancellation (SIC) are applied at the transmitter and receiver, respectively~\cite{Ding2017Application,Liu2018Multiple}. Specifically, users who have better channel conditions can remove the intra-channel interference. In the current literature, the performance gains of NOMA over orthogonal multiple access (OMA) have been investigated in various scenarios, such as cognitive radio~\cite{Lv2018Cognitive} and millimeter wave communication~\cite{Ding2017Random}. Sparked by the aforementioned benefits of the IRS and NOMA, we explore in this paper the potential performance improvement brought by \textcolor{black}{effectively integrating} NOMA technology with IRS-aided communications.
\subsection{Prior Works}
\subsubsection{Studies on Single-Input Single-Output (SISO)-NOMA Systems} Prior works on NOMA in the single-antenna scenario have studied various aspects such as user fairness, user grouping schemes, power allocation design. For example, Timotheou {\em et al.}~\cite{Timotheou2015} studied the power allocation problem with the aim of guaranteeing fairness among served users. Ding {\em et al.}~\cite{Ding2016} further analyzed the performance of NOMA in both a fixed power allocation system (F-NOMA) and cognitive radio inspired NOMA (CR-NOMA) under different user pairing schemes. Choi {\em et al.}~\cite{Choi2016} proposed optimal power allocation schemes in a two-user downlink NOMA system, where the max-sum rate and max-min rate with proportional fairness objective functions were considered. A dynamic power allocation scheme (D-NOMA) for both downlink and uplink transmission was proposed by Yang {\em et al.}~\cite{Yang2016General} while satisfying different users' quality of service (QoS) requirements, which demonstrated that D-NOMA outperforms F-NOMA and CR-NOMA in terms of communication rate and fairness. Moreover, Liu {\em et al.}~\cite{Liu2016Cooperative} studied cooperative NOMA with simultaneous wireless information and power transfer (SWIPT) technology, where near users with energy harvesting act as relays to enhance the received signal quality of far users. Fang {\em et al.}~\cite{Fang2016} studied multi-subcarrier downlink NOMA to maximize the EE by jointly optimizing subchannel assignment and power allocation, where a low complexity algorithm was designed based on matching theory and DC programming method. Full-duplex multi-subcarrier systems were investigated by Sun {\em et al.}~\cite{Sun2017Optimal}, where the optimal power allocation and user scheduling scheme was designed by applying monotonic optimization theory.
\subsubsection{Studies on Multiple-Input Multiple-Output (MIMO)-NOMA Systems} Hanif {\em et al.}~\cite{Hanif} proposed an effective beamforming design algorithm to maximize the system sum rate, where a multi-antenna base station (BS) served multiple single-antenna users through the NOMA protocol. Considering multiple antenna techniques at both the BS and users, Ding {\em et al.}~\cite{Ding2016Application} studied the precoding and detection designs, where users are partitioned into several clusters and NOMA transmission was applied in each cluster. User fairness in MIMO-NOMA systems was investigated by Liu {\em et al.}~\cite{Liu2016Fairness}, where three user clustering schemes with power allocation designs were proposed to guarantee user fairness with a lower computation complexity. Furthermore, a general framework for MIMO-NOMA in both downlink and uplink transmission was proposed in~\cite{Ding2016General}. By adopting the signal alignment technique, the MIMO-NOMA transmission can be divided into several independent single-antenna cases for NOMA implementations. Ali {\em et al.}~\cite{Ali2017Access} optimized user scheduling, beamforming vectors and power allocation in multiuser MIMO-NOMA networks, where zero-forcing beamforming was designed with the equivalent channel power gain of each cluster in order to cancel the inter-cluster interference. To investigate secure NOMA transmission, Liu {\em et al.}~\cite{Liu2017Enhancing} analyzed the secrecy performance with stochastic geometry in both single-antenna and multiple-antenna scenarios. Particularly, artificial noise was invoked to enhance secrecy performance in multiple-antenna NOMA communication. Alavi {\em et al.}~\cite{Alavi} investigated beamforming design with different objective functions based on perfect and imperfect channel state information (CSI). To further investigate the application of MIMO-NOMA, Ding {\em et al.}~\cite{Ding2016Small} proposed a precoding design scheme for Internet of Things (IoT) transmission scenarios. Wang {\em et al.}~\cite{Wang2017Spectrum} applied NOMA to millimeter-wave communication with the concept of beamspace MIMO, which demonstrated that NOMA can achieve higher system SE and EE than the conventional beamspace MIMO communication.
\subsubsection{Studies on IRS-aided Systems} In contrast to conventional communication systems, the channel response can be modified by deploying an IRS. Driven by this unique characteristic, some initial studies showed how to enhance system performance by designing the passive beamforming at the IRS. Wu {\em et al.}~\cite{Wu2019IRS} developed alternating algorithms for beanmforming design at both the BS and IRS to minimize the total transmit power, where the passive beamforming was designed with the semidefinite relaxation (SDR) approach. \textcolor{black}{To maximize the EE in the IRS-assisted multi-user multiple-input single-output (MISO) system, Huang {\em et al.}~\cite{Huang_EE} proposed two efficient algorithms by jointly optimizing the power allocation and IRS phase shifts while considering a realistic IRS power consumption model. Yu {\em et al.}~\cite{Yu_MISO} maximized the SE for an IRS-aided point-to-point MISO transmission, where the non-convex unit modulus constraint at the IRS was handled by the fixed point iteration and manifold optimization methods. Taha {\em et al.}~\cite{Taha_CSI} proposed a novel IRS architecture with several RF chains connected to the IRS controller. Based on this, two channel estimation methods were developed by utilizing compressive sensing and deep learning tools.} Chen~\cite{Chen2019Security} invoked IRSs for secure transmission in a downlink MISO system coexisting with multiple eavesdroppers. Cui {\em et al.}~\cite{Cui2019Secure} further revealed that secure transmission can still be achieved with IRSs even when the eavesdropping channel is stronger than the legitimate channel. \textcolor{black}{With the target of maximizing the sum rate of the IRS-aided system, the joint optimization of active and passive beamforming was solved by utilizing the deep reinforcement learning in~\cite{Huang_IRS_DL}.} A novel IRS-aided NOMA communication model was proposed in \cite{Ding_ax}, where IRSs were deployed at cell edge regions to maximize the total number of served users. Yang {\em et al.}~\cite{Yang_ax} investigated the max-min rate problem in both SISO and MISO IRS-aided NOMA systems. Fu {\em et al.}~\cite{Fu_ax} studied the total transmit power minimization problem in the downlink MISO NOMA IRS-aided system, where a penalty-based iterative algorithm was proposed to optimize the passive beamforming vector. \textcolor{black}{For practical implementation, IRS elements with finite resolution phase shifters were considered in~\cite{Huang_LOW,Han,Wu_Discrete,Guo_ax,You_CSI}.} \textcolor{black}{Huang {\em et al.}~\cite{Huang_LOW} studied the EE maximization problem with low resolution IRS elements by invoking the quantization method. The minimum requirement on the number of resolution bits of the IRS was derived by Han {\em et al.}~\cite{Han} to achieve an acceptable ergodic SE performance.} \textcolor{black}{A successive refinement algorithm was proposed by Wu {\em et al.}~\cite{Wu_Discrete} for the discrete IRS phase shifts design. Guo {\em et al.}~\cite{Guo_ax} investigated the weighted sum-rate maximization problem in the IRS-assisted multi-user MISO system, where three efficient algorithms were developed under different IRS element assumptions.} \textcolor{black}{You {\em et al.}~\cite{You_CSI} jointly considered the channel estimation and passive beamforming design with discrete phase shifts at the IRS.}
\vspace{-0.4cm}
\subsection{Motivation and Contributions}
It is known that the success of SIC based detection at the users in NOMA transmission is mainly determined by the channel power gains of different users. However, in IRS-aided systems, the channel response can be artificially modified by adjusting the reflection coefficients, which presents new challenges to the application of NOMA. Although few works investigated IRS-aided NOMA systems\cite{Ding_ax,Yang_ax,Fu_ax}, to the best of our knowledge, there is no existing work on the sum rate optimization problem in the MISO IRS-aided NOMA system. \textcolor{black}{In the aforementioned research contributions~\cite{Wu2019IRS,Yang_ax}, the SDR approach has been invoked for passive beamforming design. However, the Gaussian randomization approach should be adopted when the obtained solution is not rank-one, which only provides an approximate solution. The weighted sum-rate maximization problem has been studied in~\cite{Guo_ax} for a multi-user IRS-assisted MISO system under different IRS constraints. However, the proposed methods cannot be directly applied in the IRS-aided NOMA system.} The main challenges of IRS-aided NOMA networks are as follows: \textcolor{black}{i) for multi-antenna NOMA transmission, the decoding order is not determined by the users' channel power gains order, since additional decoding rate conditions need to be satisfied to guarantee successful SIC~\cite{Liu2018Multiple};} ii) both the active and passive beamforming in IRS-aided NOMA affect the decoding order among users, which makes the decoding order design and beamforming design highly coupled.\\
\indent Driven by the above challenges, in this article, we investigate the joint beamforming design at both the BS and IRS to maximize the sum rate in downlink MISO IRS-aided NOMA systems. Our main contributions of this paper are summarized as follows:
\begin{itemize}
  \item We propose a downlink MISO IRS-aided NOMA framework, \textcolor{black}{in which the IRS is utilized for SE enhancement.} Based on the proposed framework, we formulate a joint active and passive beamforming design problem to maximize the system sum rate, subject to total transmit power, SIC decoding rate constraints, user rate fairness constraints and various constraints on IRS reflecting elements, \textcolor{black}{which are defined as ideal and non-ideal IRS.}
  \item \textcolor{black}{For the ideal IRS, both the reflection amplitudes and phase shifts can be designed.} We develop an iterative algorithm using alternating optimization (AO), where the non-convex active and passive beamforming design subproblems are alternately solved by utilizing the successive convex approximation (SCA) technique. In addition, we prove that a rank-one solution can be always obtained for active beamforming design.
  \item \textcolor{black}{For the non-ideal IRS, the reflection amplitudes are fixed with constant values and only phase shifts can be designed, including continuous phase shifts and discrete phase shifts.} We first invoke a novel sequential rank-one constraint relaxation (SROCR) approach to deal with passive beamforming design problem with continuous phase shifts. In contrast to the SDR approach, the proposed algorithm can find a locally optimal rank-one solution. \textcolor{black}{Then, for discrete phase shifts, the passive beamforming is designed by leveraging the quantization method with some modifications.}
  \item \textcolor{black}{We show that the proposed algorithms are capable of achieving promising sum rate gains, compared to both the conventional system without the IRS and the IRS-aided OMA system. We also demonstrate that the performance gap between the 3-bit phase shifters and the ideal IRS is negligible.}
\end{itemize}
\vspace{-0.4cm}
\subsection{Organization and Notations}
The rest of this paper is organized as follows. Section II presents the system model and problem formulation. In Section III and Section IV, we propose efficient algorithms for the active and passive beamforming design under ideal and non-ideal IRS assumptions. Section V presents the numerical results to validate the effectiveness of the proposed designs. Finally, Section VI concludes the paper. \\
\indent \emph{Notations:} Scalars, vectors and matrices are denoted by lower-case letters, bold-face lower-case and upper-case letters, respectively. Real-valued and complex-valued matrices with the space of $N \times M$ are denoted by ${\mathbb{R}^{N \times M}}$ and ${\mathbb{C}^{N \times M}}$, respectively. All $N$-dimensional complex Hermitian matrices are denoted by ${\mathbb{H}^{N}}$. All $N$-dimensional real symmetric matrices are denoted by ${\mathbb{S}^{N}}$. ${\mathbf{I}_N}$ represents an $N \times N$ identity matrix. The rank and the trace of matrix $\mathbf{A}$ are denoted by ${\rm {rank}}\left( \mathbf{A} \right)$ and ${\rm {Tr}}\left( \mathbf{A} \right)$. ${{\mathbf{A}}} \succeq 0$ represents $\mathbf{A}$ is a positive semidefinite matrix. ${{\mathbf{a}}^T}$, ${{\mathbf{a}}^H}$ and ${\rm {diag}}\left( \mathbf{a} \right)$ denote the transpose, the conjugate transpose and the diagonal matrix of vector ${\bf{a}}$, respectively. $\operatorname{Re} \left(  \cdot  \right)$ extracts the real value of a complex variable.
\vspace{-0.4cm}
\section{System Model and Problem Formulation}
\vspace{-0.4cm}
\subsection{System Model}
\begin{figure}[h!]
    \begin{center}
        \includegraphics[width=3in]{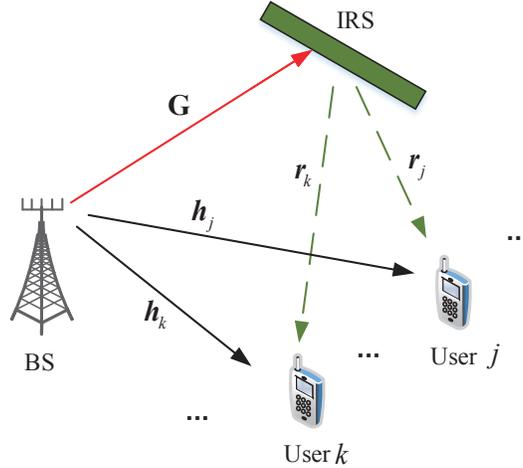}
        \caption{Illustration of the downlink MISO IRS-aided NOMA system.}
        \label{System model}
    \end{center}
\end{figure}
\vspace{-1cm}
As illustrated in Fig. \ref{System model}, we consider a downlink MISO IRS-aided NOMA system, which consists of an $N$-antenna base station, $K$ single-antenna users and an IRS equipped with $M$ passive reflecting elements. In practice, the IRS is managed by the BS through a smart controller which exchanges information and coordinates transmission. \textcolor{black}{Due to the ``double fading'' effect~\cite{Griffin}, the reflection link suffers from more severe path loss than the direct link. The power of the signals reflected by the IRS two or more times is much smaller than that of signals reflected one time and can be ignored~\cite{Wu2019IRS,Guo_ax}.} The CSI of all channels are assumed to be perfectly known at the BS\footnote{\textcolor{black}{The involved CSI can be efficiently obtained by the recently proposed channel estimation techniques for IRS-aided systems~\cite{Taha_CSI,You_CSI}. The results in this work actually provide a theoretical performance upper bound for the considered system. Our future work will relax this perfect CSI assumption by considering robust beamforming design.}} and the quasi-static flat-fading model is considered\footnote{\textcolor{black}{In this paper, the users are assumed to be static or moving at a low speed in the considered networks. As a result, the channels would remain unchanged for a long time duration and the time consumption for channel acquisitions can be ignored.}}. Let $\Theta  = {\rm {diag}}\left( {\mathbf{u}} \right) \in {\mathbb{C}^{M \times M}}$ denote the diagonal reflection coefficients matrix of the IRS with ${\mathbf{u}} = \left[ {{u_1},{u_2}, \cdots ,{u_M}} \right]$ and
${u_m} = {\beta _m}{e^{j{\theta _m}}}$, where ${\beta _m} \in \left[ {0,1} \right]$ and ${\theta _m} \in \left[ {0,2\pi } \right)$ denote the reflection amplitude and phase shift of the $m$th reflecting element on the IRS, respectively. Depending on the amplitude and phase shift features of reflecting elements, two sets of IRS assumptions are considered as follows~\cite{IRS_ax}:
\begin{itemize}
  \item \textbf{Ideal IRS}: In this scenario, the reflecting elements can be optimized with arbitrary continuous amplitudes and phase shifts. Thus, the feasible set of ${u_m}$ can be expressed as
      \vspace{-0.4cm}
      \begin{align}\label{phi 1}
      {\Phi _1} = \left\{ {{u_m}|{{\left| {{u_m}} \right|}^2} \in \left[ {0,1} \right]} \right\}.
      \end{align}
      \vspace{-1.2cm}
  \item \textbf{Non-ideal IRS}: In this scenario, the reflection amplitude is fixed with a constant value, such as ${\beta _m} = 1$. We further discuss the two scenarios of continuous phase shifts and discrete phase shifts. In particular, the feasible set of ${u_m}$ for continuous phase shifts can be expressed as
      \vspace{-0.6cm}
      \begin{align}\label{phi 2}
      {\Phi _2} = \left\{ {{u_m}|{u_m} = {e^{j{\theta _m}}},{\theta _m} \in \left[ {0,2\pi } \right)} \right\}.
      \end{align}
      \vspace{-1.2cm}

      \noindent and the feasible set of ${u_m}$ for discrete phase shifts with $B$ resolution bits can be expressed as
      \vspace{-0.6cm}
      \begin{align}\label{phi 3}
      {\Phi _3} = \left\{ {{u_m}|{{\left| {{u_m}} \right|}^2} = 1,{\theta _m} \in {\mathcal{D}}} \right\},
      \end{align}
      \vspace{-1.2cm}

      \noindent where ${\mathcal{D}} = \left\{ {\frac{{n2\pi }}{{{2^B}}},n = 0,1,2, \cdots ,{2^B} - 1} \right\}$.
\end{itemize}
The channels from the BS to user $k$, the IRS to user $k$ and the BS to IRS are denoted by ${{\mathbf{h}}_k} \in {\mathbb{C}^{N \times 1}}$, ${{\mathbf{r}}_k} \in {\mathbb{C}^{M \times 1}}$ and ${\mathbf{G}} \in {\mathbb{C}^{M \times N}}$, respectively. Let $s_k$ denote the information-bearing symbol for the $k$th user with zero mean and unit variance. Therefore, the complex baseband signal transmitted from the BS can be expressed as ${\mathbf{x}} = \sum\limits_{k = 1}^K {{{\mathbf{w}} _k}{s_k}} $, where ${{\mathbf{w}} _k} \in {\mathbb{C}^{N \times 1}}$ represents the active beamforming vector for the $k$th user. Then, the received signal at user $k$ can be expressed as
\vspace{-0.6cm}
\begin{align}\label{received signal}
{y_k} = \left( {{\mathbf{h}}_k^H + {\mathbf{r}}_k^H\Theta {\mathbf{G}}} \right)\sum\limits_{k = 1}^K {{{\mathbf{w}} _k}{s_k}}  + {n_k},
\end{align}
\vspace{-0.8cm}

\noindent where ${n_k}\sim\mathcal{C}\mathcal{N}\left( {0,{\sigma ^2}} \right)$ is the additive white Gaussian noise (AWGN) at user $k$ with zero mean and variance ${\sigma ^2}$.\\
\indent Based on the NOMA principle, each user tries to employ SIC to remove the intra-cell interference. In SISO NOMA systems, the optimal decoding order among users are determined by the channel power gains. The user with a stronger channel power gains can decode the user's signal who has a weaker channel power gain. However, this ordering method cannot be applied in a MISO NOMA system since the channel responses can be modified with the introduced IRS, which can change the decoding order to be any one of all the $K!$ different decoding orders. Let $\Omega \left( k \right)$ denote the decoding order of user $k$. For instance, if $\Omega \left( k \right) = k$, then user $k$ is the $k$th signal to be decoded. \textcolor{black}{Based on this order, user $k$ first successively decodes the signal of each user $m$ with $\Omega \left( m \right) < \Omega \left( k \right)$ before decoding its own signal, and treats the signal of each user $i$ with $\Omega \left( i \right) > \Omega \left( k \right)$ as interference~\cite{Liu2017,Ding2017Application}. Compared with the conventional communication system, the implementation of NOMA technology imposes additional complexity since the users have to decode information intended for some of the other users.} \textcolor{black}{Therefore, the achievable signal-to-interference-plus-noise ratio (SINR) at user $k$ to decode its own signal can be expressed as}
\vspace{-0.6cm}
\begin{align}\label{SINR k k}
\textcolor{black}{{\rm{SINR}}_{k \to k}^{} = \frac{{{{\left| {\left( {{\mathbf{h}}_k^H + {\mathbf{r}}_k^H\Theta {\mathbf{G}}} \right){{\mathbf{w}} _k}} \right|}^2}}}{{\sum\limits_{\Omega \left( i \right) > \Omega \left( k \right)} {{{\left| {\left( {{\mathbf{h}}_k^H + {\mathbf{r}}_k^H\Theta {\mathbf{G}}} \right){{\mathbf{w}} _i}} \right|}^2}}  + {\sigma ^2}}}.}
\end{align}
\vspace{-0.8cm}

\noindent \textcolor{black}{The corresponding achievable rate at user $k$ to decode it own signal is ${R_{k \to k}}\! =\! {\log _2}\left( {1\! +\! {\rm{SINR}}_{k \to k}} \right)$.}\\
\textcolor{black}{\indent Furthermore, for any users $k$ and $j$ which satisfy $\Omega \left( k \right) < \Omega \left( j \right)$, the SINR at user $j$ to decode user $k$'s signal can be expressed as}
\vspace{-0.3cm}
\begin{align}\label{SINR k j}
\textcolor{black}{{\rm{SINR}}_{k \to j}^{} = \frac{{{{\left| {\left( {{\mathbf{h}}_j^H + {\mathbf{r}}_j^H\Theta {\mathbf{G}}} \right){{\mathbf{w}} _k}} \right|}^2}}}{{\sum\limits_{\Omega \left( i \right) > \Omega \left( k \right)} {{{\left| {\left( {{\mathbf{h}}_j^H + {\mathbf{r}}_j^H\Theta {\mathbf{G}}} \right){{\mathbf{w}} _i}} \right|}^2}}  + {\sigma ^2}}}.}
\end{align}
\vspace{-0.8cm}

\noindent \textcolor{black}{The corresponding achievable rate at user $j$ to decode user $k$'s signal is ${R_{k \to j}}\! =\! {\log _2}\left( {1\! +\! {\rm{SINR}}_{k \to j}} \right)$.}\\
\indent \textcolor{black}{It is worth noting that in order to guarantee that the SIC can be performed successfully, the achievable rate at user $j$ to decode user $k$'s signal should be no less than the achievable rate at user $k$ to decode it own signal~\cite{Liu2018Multiple}. Then, we have the following SIC decoding rate constraints}
\vspace{-0.3cm}
\begin{align}\label{SIC decoding constraints}
\textcolor{black}{{R_{k \to j}} \ge {R_{k \to k}}, \Omega \left( j \right) > \Omega \left( k \right).}
\end{align}
\vspace{-1cm}

\noindent \textcolor{black}{In addition, to guarantee rate fairness among all users, the following conditions should be satisfied with given decoding orders}
\vspace{-0.3cm}
\begin{align}\label{fairness}
\textcolor{black}{
  {\left| {\left( {{\mathbf{h}}_k^H + {\mathbf{r}}_k^H\Theta {\mathbf{G}}} \right){{\mathbf{w}}_{\Omega \left( i \right)}}} \right|^2} \le {\left| {\left( {{\mathbf{h}}_k^H + {\mathbf{r}}_k^H\Theta {\mathbf{G}}} \right){{\mathbf{w}}_{\Omega \left( j \right)}}} \right|^2}, \forall k,i,j \in {\mathcal{K}},\Omega \left( i \right) > \Omega \left( j \right).
}
\end{align}
\vspace{-1cm}

\noindent \textcolor{black}{From equation \eqref{SINR k k}, it can be observed that users with higher decoding orders receive less or no interference from other users due to the SIC. The inequalities in \eqref{fairness} avoid the case that most of the radio resources are allocated to some of the higher decoding order users. With the above inequalities, the received signal power of users in lower decoding orders is made greater than that of users with higher decoding orders. As a result, a reasonable communication rate can be achieved at users in lower decoding orders, which is capable of maintaining the rate fairness among users.}\\
\indent For example, let us consider the two-user case. If the decoding order is set as $\Omega \left( k \right) = k,k = 1,2$, then SIC decoding and rate fairness conditions can be expressed as
\vspace{-0.4cm}
\begin{subequations}
\begin{align}
\label{SIC 2user}&{R_{1 \to 2}} \ge {R_{1 \to 1}},\\
\label{fairness 2user}&{\left| {\left( {{\mathbf{h}}_k^H + {\mathbf{r}}_k^H\Theta {\mathbf{G}}} \right){{\mathbf{w}}_2}} \right|^2} \le {\left| {\left( {{\mathbf{h}}_k^H + {\mathbf{r}}_k^H\Theta {\mathbf{G}}} \right){{\mathbf{w}}_1}} \right|^2},k = 1,2.
\end{align}
\end{subequations}
\vspace{-0.8cm}

\noindent For the three-user case with the decoding order $\Omega \left( k \right) = k,k = 1,2,3$, the SIC decoding rate conditions at user 2 and user 3 and rate fairness condition among all users can be expressed as
\vspace{-0.4cm}
\begin{subequations}
\begin{align}
\label{SIC 3user}&{R_{1 \to 2}} \ge {R_{1 \to 1}},\;{R_{1 \to 3}} \ge {R_{1 \to 1}},\;\;{R_{2 \to 3}} \ge {R_{2 \to 2}},\\
\label{fairness 3user}&
  {\left| {\left( {{\mathbf{h}}_k^H + {\mathbf{r}}_k^H\Theta {\mathbf{G}}} \right){{\mathbf{w}}_3}} \right|^2} \le {\left| {\left( {{\mathbf{h}}_k^H + {\mathbf{r}}_k^H\Theta {\mathbf{G}}} \right){{\mathbf{w}}_2}} \right|^2} \le {\left| {\left( {{\mathbf{h}}_k^H + {\mathbf{r}}_k^H\Theta {\mathbf{G}}} \right){{\mathbf{w}}_1}} \right|^2},k = 1,2,3.
\end{align}
\end{subequations}
\vspace{-0.8cm}

\noindent It is worth noting that when $K$ users are served in the IRS-aided NOMA system, there will be $\frac{{K\left( {K - 1} \right)}}{2}$ SIC decoding rate constraints and $K\left( {K - 1} \right)$ rate fairness constraints, which depend on not only the active beamforming coefficients ${\left\{ {{{\mathbf{w}}_k}} \right\}}$ at the BS, but also the combined channel (which depends on the passive beamforming vector at the IRS).
\vspace{-0.4cm}
\subsection{Problem Formulation}
Our goal is to maximize the sum rate of all users by jointly optimizing the active beamforming coefficients ${\left\{ {{{\mathbf{w}} _k}} \right\}}$ at the BS and the passive beamforming vector at the IRS, subject to the total power constraint, the SIC decoding rate and the user rate fairness constraints under different IRS assumptions. The optimization problem can be formulated as
\begin{subequations}
\begin{align}\label{P1}
({{\rm{P1}}}):&\mathop {\max }\limits_{\Omega ,\Theta ,\left\{ {{{\mathbf{w}}_k}} \right\}} \;\;\sum\limits_{k = 1}^K {{R_{k \to k}}}  \\
\label{SIC}{\rm{s.t.}}\;\;&{R_{k \to j}} \ge {R_{k \to k}},\;\Omega \left( j \right) > \Omega \left( k \right),\\
\label{total power}&\sum\limits_{k = 1}^K {\left\| {{{\mathbf{w}}_k}} \right\|}^2  \le {P_T}\\
\label{discrete phase shift}&{u _m} \in \Phi,\\
\label{decoding order}&\Omega  \in \Pi ,\\
\label{constraint P1}&\eqref{fairness}.
\end{align}
\end{subequations}
where $\Pi$ denotes the set of all $K!$ possible SIC decoding orders and $P_T$ denotes the total transmit power. Constraint \eqref{SIC} guarantees that the SIC can be performed successfully. Constraint \eqref{total power} is the total transmission power constraint and \eqref{discrete phase shift} represents the considered IRS assumption. Constraint \eqref{constraint P1} guarantees rate fairness among all users. \textcolor{black}{It is worth noting that Problem (P1) is always feasible under any given $P_T$, since we consider the rate fairness constraints instead of the strict QoS constraints~\cite{Wu2019IRS,Huang_EE}.} However, Problem (P1) is a highly-coupled non-convex problem even with convex set ${\Phi _1}$, which makes it hard to find the global optimal solution. In the following, efficient algorithms are developed based on the AO method to derive a high quality suboptimal solution.
\vspace{-0.4cm}
\section{Ideal IRS Case}
In this section, we focus on the Problem (P1) with the ideal IRS. Before solving this problem, we first transform (P1) into a more tractable form. Let ${{{\mathbf{H}}}_k} = \left[ \begin{gathered}
  {\rm {diag}}\left( {{\mathbf{r}}_k^H} \right){\mathbf{G}} \hfill \\
  {\mathbf{h}}_k^H \hfill \\
\end{gathered}  \right]$ and ${\mathbf{v}} = {\left[ {{\mathbf{u}}\;1} \right]^H}$. Then, we have ${\left| {\left( {{\mathbf{h}}_k^H + {\mathbf{r}}_k^H\Theta {\mathbf{G}}} \right){{\mathbf{w}} _k}} \right|^2} = {\left| {{{\mathbf{v}}^H}{{\mathbf{H}}_k}{{\mathbf{w}} _k}} \right|^2}$. Furthermore, we introduce slack variables $\left\{ {{S_{kj}}} \right\}$ and $\left\{ {{I_{kj}}} \right\}$ such that
\vspace{-0.4cm}
\begin{align}\label{S_kj}
\frac{1}{{{S_{kj}}}} = {\left| {{{\mathbf{v}}^H}{{\mathbf{H}}_j}{{\mathbf{w}}_k}} \right|^2},
\end{align}
\vspace{-1.2cm}
\begin{align}\label{I_kj}
{I_{kj}} = \sum\limits_{\Omega \left( i \right) > \Omega \left( k \right)} {{{\left| {{{\mathbf{v}}^H}{{\mathbf{H}}_j}{{\mathbf{w}}_i}} \right|}^2}}  + {\sigma ^2},
\end{align}
\vspace{-0.8cm}

\noindent Then, the decoding rate ${R_{k \to j}}$ can be expressed as
\begin{align}\label{R_jk}
{R_{k \to j}} = {\log _2}\left( {1 + \frac{1}{{{S_{kj}}{I_{kj}}}}} \right),\Omega \left( k \right) \le \Omega \left( j \right).
\end{align}
With the above variable definitions, Problem (P1) can be transformed into
\begin{subequations}
\begin{align}\label{P2}
({{\rm{P2}}}):&\mathop {\max }\limits_{{\mathbf{v}},\left\{ {{{\mathbf{w}}_k},{R_{k \to j}},{S_{kj}},{I_{kj}}} \right\}} \;\;\sum\limits_{k = 1}^K {{R_{k \to k}}}     \\
\label{R_kk P2}{\rm{s.t.}}\;\;&{R_{k \to j}} \le {\log _2}\left( {1 + \frac{1}{{{S_{kj}}{I_{kj}}}}} \right),\Omega \left( k \right) \le \Omega \left( j \right), \forall k,j \in {\mathcal{K}}\\
\label{S_kj P2}&\frac{1}{{{S_{kj}}}} \le {\left| {{{\mathbf{v}}^H}{{\mathbf{H}}_j}{{\mathbf{w}}_k}} \right|^2},\forall k,j \in {\mathcal{K}},\\
\label{I_kj P2}&{I_{kj}} \ge \sum\limits_{\Omega \left( i \right) > \Omega \left( k \right)} {{{\left| {{{\mathbf{v}}^H}{{\mathbf{H}}_j}{{\mathbf{w}}_i}} \right|}^2}}  + {\sigma ^2},\forall k,j \in {\mathcal{K}},\\
\label{SIC P2}&{R_{k \to j}} \ge {\log _2}\left( {1 + \frac{1}{{{S_{kk}}{I_{kk}}}}} \right),\Omega \left( j \right) > \Omega \left( k \right),\forall k,j \in {\mathcal{K}},\\
\label{fairness P2}&{\left| {{{\mathbf{v}}^H}{{\mathbf{H}}_j}{{\mathbf{w}}_k}} \right|^2} \ge {\left| {{{\mathbf{v}}^H}{{\mathbf{H}}_j}{{\mathbf{w}}_i}} \right|^2},\Omega \left( k \right) < \Omega \left( i \right),\forall i,j,k,\\
\label{constraints P2}&\eqref{total power}-\eqref{decoding order}.
\end{align}
\end{subequations}
\begin{proposition}\label{P2 P1}
\emph{Problem (P2) is equivalent to Problem (P1).}
\begin{proof}
\emph{Without loss of optimality to Problem (P1), constraints \eqref{R_kk P2}, \eqref{S_kj P2} and \eqref{I_kj P2} can be met with equality. Specifically, when $k=j$, assume that if any of the constraints in \eqref{R_kk P2} is satisfied with strict inequality, then we can always increase ${R_{k \to j}}$ to make the constraint \eqref{R_kk P2} satisfied with equality while increasing the objective function's value. Furthermore, suppose that \eqref{S_kj P2} and \eqref{I_kj P2} are satisfied with strict inequality, then we can always reduce ${S_{kj}}$ or increase ${I_{kj}}$ to make all constraints satisfied with equality, which in turn increases the value of the right-hand-side (RHS) in \eqref{R_kk P2} and increases the objective function's value. When $k \ne j$, assume that if any of the constraints in \eqref{R_kk P2} is satisfied with strict inequality, then we can always increase ${R_{k \to j}}$ without changing the objective function's value of (P1). Therefore, Problem (P2) is equivalent to Problem (P1).}
\end{proof}
\end{proposition}
However, Problem (P2) is still a non-convex optimization problem since the decoding order $\Omega $, the active beamforming coefficients $\left\{ {{{\mathbf{w}}_k}} \right\}$ and the passive beamforming vector ${\mathbf{v}}$ are highly-coupled. Since the total number of decoding order combinations is a finite value, the optimal sum rate can be obtained by solving Problem (P2) with any one of decoding orders at first and selecting the maximum objective function's value among all decoding orders. For a given decoding order, the sum rate maximization problem in (P2) with the ideal IRS is reduced to
\vspace{-0.3cm}
\begin{subequations}
\begin{align}\label{P3}
({{\rm{P3}}}):&\mathop {\max }\limits_{{\mathbf{v}},\left\{ {{{\mathbf{w}}_k},{R_{k \to j}},{S_{kj}},{I_{kj}}} \right\}} \;\;\sum\limits_{k = 1}^K {{R_{k \to k}}}     \\
\label{discrete phase shift P3}{\rm{s.t.}}\;\;&0 \le \left| {{v_m}} \right|^2 \le 1,m = 1,2, \cdots ,M,{v_{M + 1}} = 1,\\\
\label{constraints P3}&\eqref{total power},\eqref{R_kk P2}-\eqref{fairness P2}.
\end{align}
\end{subequations}
\vspace{-1cm}

\noindent  In order to tackle the highly-coupled non-convex terms in Problem (P3), we decompose the original problem into the two subproblems of active beamforming optimization and passive beamforming optimization, which can be efficiently solved by the SCA technique as described next.
\vspace{-0.4cm}
\subsection{Active Beamforming Optimization}
Define ${{\mathbf{W}}_k} = {{\mathbf{w}} _k}{\mathbf{w}} _k^H,\forall k$, which satisfy ${{\mathbf{W}}_k} \succeq 0$ and ${\rm {rank}}\left( {{{\mathbf{W}}_k}} \right) = 1$. Under any given feasible passive beamforming vector ${\mathbf{v}}$, the active beamforming optimization problem can be written as
\begin{subequations}
\begin{align}\label{P3.1}
({{\rm{P3.1}}}):&\mathop {\max }\limits_{\left\{ {{{\mathbf{W}}_k},{R_{k \to j}},{S_{kj}},{I_{kj}}} \right\}} \;\;\sum\limits_{k = 1}^K {{R_{k \to k}}}     \\
\label{S_kj P3.1}{\rm{s.t.}}\;\;&\frac{1}{{{S_{kj}}}} \le {\rm{Tr}}\left( {{{\mathbf{W}}_k}{\mathbf{H}}_j^H{\mathbf{v}}{{\mathbf{v}}^H}{{\mathbf{H}}_j}} \right),\forall k,j \in {{\mathcal{K}}},\\
\label{I_kj P3.1}&{I_{kj}} \ge \sum\limits_{\Omega \left( i \right) > \Omega \left( k \right)} {{\rm{Tr}}\left( {{{\mathbf{W}}_i}{\mathbf{H}}_j^H{\mathbf{v}}{{\mathbf{v}}^H}{{\mathbf{H}}_j}} \right)}  + {\sigma ^2},\forall k,j \in {{\mathcal{K}}},\\
\label{fairness P3.1}&
  {\rm{Tr}}\left( {{{\mathbf{W}}_k}{\mathbf{H}}_j^H{\mathbf{v}}{{\mathbf{v}}^H}{{\mathbf{H}}_j}} \right) \ge {\rm{Tr}}\left( {{{\mathbf{W}}_i}{\mathbf{H}}_j^H{\mathbf{v}}{{\mathbf{v}}^H}{{\mathbf{H}}_j}} \right), \Omega \left( k \right) < \Omega \left( i \right),\forall i,j,k, \\
\label{total power W}&\sum\limits_{k = 1}^K {{\rm{Tr}}\left( {{{\mathbf{W}}_k}} \right)}  \le {P_T},\\
\label{SDP W}&{{\mathbf{W}}_k}  \succeq  0, {\mathbf{W}}_k \in {{\mathbb{H}}^{N}}, \forall k,\\
\label{rank 1 W}&{\rm {rank}}\left( {{{\mathbf{W}}_k}} \right) = 1,\forall k,\\
\label{constraints P3.1}&\eqref{R_kk P2},\eqref{SIC P2}.
\end{align}
\end{subequations}
\begin{lemma}\label{lemma:xy}
\emph{For $x > 0$ and $y > 0$, $f\left( {x,y} \right) = {\log _2}\left( {1 + \frac{1}{{xy}}} \right)$ is a convex function with respect to $x$ and $y$.}
\begin{proof}
\emph{It is easy to prove lemma \ref{lemma:xy} by showing that the Hessian matrix of function $f\left( {x,y} \right)$ is positive semidefinite when $x > 0$ and $y > 0$. As a result, $f\left( {x,y} \right)$ is a convex function.}
\end{proof}
\end{lemma}
Problem (P3.1) is a non-convex problem due to the non-convex constraints \eqref{R_kk P2} and \eqref{rank 1 W}. Based on \textbf{Lemma \ref{lemma:xy}}, the RHS of \eqref{R_kk P2} is a joint convex function with respect to ${{S_{kj}}}$ and ${{I_{kj}}}$. By applying the first-order Taylor expansion, the lower bound at given local points $\left\{ {{S_{kj}^l},{I_{kj}^l}} \right\}$ can be expressed as
\begin{align}\label{Rjk lower bound}
   {\log _2}\left( {1 + \frac{1}{{{S_{kj}}{I_{kj}}}}} \right) \ge R_{kj}^{low} = \log_2 \left( {1 + \frac{1}{{S_{kj}^lI_{kj}^l}}} \right) - \frac{{\left( {{{\log }_2}e} \right)\left( {S_{kj} - S_{kj}^l} \right)}}{{S_{kj}^l + S_{kj}^l{^2}I_{kj}^l}} - \frac{{\left( {{{\log }_2}e} \right)\left( {I_{kj}^{} - I_{kj}^l} \right)}}{{I_{kj}^l + I_{kj}^l{{}^2}S_{kj}^l}}.
\end{align}
Then, Problem (P3.1) is approximated as the following problem
\vspace{-0.3cm}
\begin{subequations}
\begin{align}\label{P3.2}
({{\rm{P3.2}}}):&\mathop {\max }\limits_{\left\{ {{{\mathbf{W}}_k},{R_{k \to j}},{S_{kj}},{I_{kj}}} \right\}} \;\;\sum\limits_{k = 1}^K {{R_{k \to k}}}     \\
\label{R_kk P3.2}{\rm{s.t.}}\;\;&{R_{k \to j}} \le R_{kj}^{low},\forall k,j \in {{\mathcal{K}}},\\
\label{constraints P3.1}&\eqref{SIC P2},\eqref{S_kj P3.1}-\eqref{rank 1 W}.
\end{align}
\end{subequations}
\vspace{-1cm}

\noindent Note that the remaining non-convexity of (P3.2) is the rank-one constraint \eqref{rank 1 W}. To tackle this issue, we have the following theorem
\begin{theorem}\label{rank one relax}
\emph{Without loss of optimality, the obtained solution to Problem (P3.2) without rank-one constraint \eqref{rank 1 W} can always satisfy ${\rm {rank}}\left( {{{\mathbf{W}}_k}} \right) \le 1,\forall k \in {{\mathcal{K}}}$. }
\begin{proof}
See Appendix A.
\end{proof}
\end{theorem}
Based on \textbf{Theorem \ref{rank one relax}}, we can always obtain a rank-one solution by solving (P3.2) by ignoring the rank-one constraint \eqref{rank 1 W}. As a result, the relaxed problem is a convex semidefinite program (SDP), which can be efficiently solved via standard convex problem solvers such as CVX~\cite{cvx}. Note that the objective function's value obtained from Problem (P3.2) in general provides a lower bound on that of Problem (P3.1) due to the replacement of the lower bounds \eqref{Rjk lower bound}. After solving (P3.2), the active beamforming coefficients $\left\{ {{{\mathbf{w}}_k}} \right\}$ can be obtained through Cholesky decomposition, e.g. ${\mathbf{W}}_k^* = {{\mathbf{w}}_k}{\mathbf{w}}_k^H,\forall k$.
\vspace{-0.4cm}
\subsection{Passive Beamforming Optimization with $\Phi_1$}
Given any feasible active beamforming vectors ${\left\{ {{{\mathbf{w}}_k}} \right\}}$, the passive beamforming optimization problem with ideal IRS can be written as
\vspace{-0.3cm}
\begin{subequations}
\begin{align}\label{P3}
({{\rm{P3.3}}}):&\mathop {\max }\limits_{{\mathbf{v}},\left\{ {{R_{k \to j}},{S_{kj}},{I_{kj}}} \right\}} \;\;\sum\limits_{k = 1}^K {{R_{k \to k}}}     \\
\label{constraints P3.3}{\rm{s.t.}}\;\;&\eqref{R_kk P2}-\eqref{fairness P2},\eqref{discrete phase shift P3}.
\end{align}
\end{subequations}
\vspace{-1cm}

\noindent Problem (P3.3) is a non-convex problem due to non-convex constraints \eqref{R_kk P2}, \eqref{S_kj P2} and \eqref{fairness P2}. In the previous subsection, we have already showed how to tackle the non-convex constraint \eqref{R_kk P2}. For the non-convex constraint \eqref{S_kj P2}, the RHS is a convex function with respect to ${\mathbf{v}}$, the lower bound with the first-order Taylor expansion at the given local point ${{\mathbf{v}}^{\left( l \right)}}$ can be expressed as
\vspace{-0.3cm}
\begin{align}\label{v lower bound}
  {\left| {{{\mathbf{v}}^H}{{\mathbf{H}}_j}{{\mathbf{w}}_k}} \right|^2} \ge {\lambda _{kj}} = {\left| {{{\mathbf{v}}^{\left( l \right)}}^H{{\mathbf{H}}_j}{{\mathbf{w}}_k}} \right|^2} + 2\operatorname{Re} \left( {\left( {{{\mathbf{v}}^{\left( l \right)}}^H{{\mathbf{H}}_j}{{\mathbf{w}}_k}{\mathbf{w}}_k^H{\mathbf{H}}_j^H} \right)\left( {{\mathbf{v}} - {{\mathbf{v}}^{\left( l \right)}}} \right)} \right).
\end{align}
\vspace{-1cm}

\noindent Similarly, for the non-convex constraint \eqref{fairness P2}, the left hand side (LHS) can be replaced with the lower bound in \eqref{v lower bound}. Then, the passive beamforming optimization problem is approximated as the following problem
\vspace{-0.3cm}
\begin{subequations}
\begin{align}\label{P3.4}
({{\rm{P3.4}}}):&\mathop {\max }\limits_{{\mathbf{v}},\left\{ {{R_{kj}},{S_{kj}},{I_{kj}}} \right\}} \;\;\sum\limits_{k = 1}^K {{R_{k \to k}}}     \\
\label{R_kk P3.4}{\rm{s.t.}}\;\;&{R_{k \to j}} \le R_{kj}^{low},\forall k,j \in {{\mathcal{K}}},\\
\label{S_kj P3.4}&\frac{1}{{{S_{kj}}}} \le {\lambda _{kj}},\\
\label{fairness P3.4}&{\lambda _{kj}} \ge {\left| {{{\mathbf{v}}^H}{{\mathbf{H}}_j}{{\mathbf{w}}_i}} \right|^2},\Omega \left( k \right) < \Omega \left( i \right),\forall i,j,k,\\
\label{constraints P3.4}&\eqref{I_kj P2},\eqref{SIC P2},\eqref{discrete phase shift P3}.
\end{align}
\end{subequations}
\vspace{-1cm}

\noindent Now, it is easy to verify that Problem (P3.4) is a convex problem, which can be efficiently solved via standard convex problem solvers such as CVX~\cite{cvx}. Similarly, the objective function's value obtained from (P3.4) serves as a lower bound on that of (P3.3).
\vspace{-0.4cm}
\subsection{Proposed Algorithm, Complexity and Convergence}
Based on the above two subproblems, we propose an iterative algorithm for Problem (P3) with the ideal IRS by utilizing the AO method. Specifically, the active beamforming coefficients $\left\{ {{{\mathbf{w}}_k}} \right\}$ and the passive beamforming vector ${\mathbf{v}}$ are alternately optimized by solving Problem (P3.2) and (P3.4), where the solutions obtained after each iteration are used as the input local points for the next iteration. The details of the proposed algorithm are summarized in \textbf{Algorithm 1}. According to \cite{Luo}, the complexity of the SDP subproblem for active beamforming design is ${\mathcal{O}}\left( {\max {{\left( {N,3K\left( {K - 1} \right)} \right)}^4}\sqrt N \log \frac{1}{\varepsilon }} \right)$, where $\varepsilon$ denotes the accuracy. The complexity of the subproblem for passive beamforming design with the interior-point method is ${\mathcal{O}}\left( {{{\left( {3{K^2} + {M}} \right)}^{3.5}}} \right)$ \textcolor{black}{~\cite{convex}}. Then, the total complexity of Algorithm 1 is ${{\mathcal{O}}}\left( {I_{ite}^I\left( {\max {{\left( {N,3K\left( {K - 1} \right)} \right)}^4}\sqrt N \log \frac{1}{\varepsilon } + } \right.} \right.$\\$\left. {\left. {{{\left( {3{K^2} + M} \right)}^{3.5}}} \right)} \right)$, where ${I_{ite}^I}$ denotes the number of iterations for Algorithm 1.\\
\begin{algorithm}[!t]\label{method1}
\caption{Proposed SCA-based algorithm for solving Problem (P3) with $\Phi_1$} 
 \hspace*{0.02in} 用来控制位置，同时利用 \\ 进行换行
\hspace*{0.02in} {Initialize a decoding order $\Omega $ and feasible solutions $\left\{ {{\mathbf{w}}_k^l} \right\},{{\mathbf{v}}^l}$ to (P3), $l=0$.}\\
\vspace{-0.4cm}
\begin{algorithmic}[1]
\STATE {\bf repeat}
\STATE Solve Problem (P3.2) for given ${\mathbf{v}}^l$, where the optimal solution is denoted by $\left\{ {{\mathbf{w}}_k^{l + 1}} \right\}$.
\STATE Solve Problem (P3.4) for given $\left\{ {{\mathbf{w}}_k^{l + 1}} \right\}$, where the optimal solution is denoted by ${\mathbf{v}}^{l+1}$.
\STATE $l=l+1$.
\STATE {\bf until} the fractional increase of the objective value is below a threshold $\xi   > 0$.
\end{algorithmic}
\end{algorithm}
\indent Next, we demonstrate the convergence of Algorithm 1. Define ${\eta _{{\Phi _1}}}\left( {\left\{ {{\mathbf{w}}_k^l} \right\},{{\mathbf{v}}^l}} \right)$ as the objective function's value of Problem (P3) in the $l$th iteration. First, for Problem (P3.2) with given passive beamforming vector in step 2 of Algorithm 1, we have
\begin{align}\label{P3.1 convergence}
  {\eta _{{\Phi _1}}}\left( {\left\{ {{\mathbf{w}}_k^l} \right\},{{\mathbf{v}}^l}} \right)\mathop  = \limits^{\left( a \right)} \eta _{\mathbf{w}}^{lb}\left( {\left\{ {{\mathbf{w}}_k^l} \right\},{{\mathbf{v}}^l}} \right) \mathop  \le \limits^{\left( b \right)} \eta _{\mathbf{w}}^{lb}\left( {\left\{ {{\mathbf{w}}_k^{l + 1}} \right\},{{\mathbf{v}}^l}} \right) \mathop  \le \limits^{\left( c \right)} {\eta _{{\Phi _1}}}\left( {\left\{ {{\mathbf{w}}_k^{l + 1}} \right\},{{\mathbf{v}}^l}} \right),
\end{align}
where $\eta _{\mathbf{w}}^{lb}$ represents the objective function's value of Problem (P3.2). ${\left( a \right)}$ follows the fact that the first-order Taylor expansions are tight at the given local points in Problem (P3.2); ${\left( b \right)}$ holds since Problem (P3.2) is solved optimally; ${\left( c \right)}$ holds due to the fact that the objective function's value of Problem (P3.2) serves as a lower bound on that of (P3.1). This suggests that the objective function's value of Problem (P3.1) is non-decreasing after each iteration.\\
\indent Similarly, for Problem (P3.4) with given active beamforming coefficients in step 3 of Algorithm 1, we have
\begin{align}\label{P3.3 convergence}
  {\eta _{{\Phi _1}}}\left( {\left\{ {{\mathbf{w}}_k^{l+1}} \right\},{{\mathbf{v}}^l}} \right)\mathop  =  \eta _{\mathbf{v}}^{lb}\left( {\left\{ {{\mathbf{w}}_k^{l+1}} \right\},{{\mathbf{v}}^l}} \right) \mathop  \le  \eta _{\mathbf{v}}^{lb}\left( {\left\{ {{\mathbf{w}}_k^{l + 1}} \right\},{{\mathbf{v}}^{l+1}}} \right) \mathop  \le  {\eta _{{\Phi _1}}}\left( {\left\{ {{\mathbf{w}}_k^{l + 1}} \right\},{{\mathbf{v}}^{l+1}}} \right),
\end{align}
where $\eta _{\mathbf{v}}^{lb}$ represents the objective function's value of Problem (P3.4).\\
\indent As a result, based on \eqref{P3.1 convergence} and \eqref{P3.3 convergence}, we obtain that
\vspace{-0.4cm}
\begin{align}\label{overall convergence P3}
\begin{gathered}
{\eta _{{\Phi _1}}}\left( {\left\{ {{\mathbf{w}}_k^l} \right\},{{\mathbf{v}}^l}} \right) \le {\eta _{{\Phi _1}}}\left( {\left\{ {{\mathbf{w}}_k^{l + 1}} \right\},{{\mathbf{v}}^{l + 1}}} \right).
\end{gathered}
\end{align}
\begin{remark}\label{convergence}
\textcolor{black}{\emph{\textcolor{black}{Equation \eqref{overall convergence P3} indicates that the objective function's value of Problem (P3) is non-decreasing after each iteration. Since the system sum rate is upper bounded by a finite value, the proposed algorithm is guaranteed to converge.}}}
\end{remark}
\vspace{-0.6cm}
\section{Non-Ideal IRS Case}
In this section, we solve Problem (P2) with non-ideal IRS. The sum rate maximization problem in (P2) for a given decoding order can be written as
\begin{subequations}
\begin{align}\label{P4}
({{\rm{P4}}}):&\mathop {\max }\limits_{{\mathbf{v}},\left\{ {{{\mathbf{w}}_k},{R_{k \to j}},{S_{kj}},{I_{kj}}} \right\}} \;\;\sum\limits_{k = 1}^K {{R_{k \to k}}}     \\
\label{discrete phase shift P4}{\rm{s.t.}}\;\;&{v_m} \in {\Phi _2}\;{\rm{or}}\;{\Phi _3},m = 1,2, \cdots ,M,{v_{M + 1}} = 1,\\\
\label{constraints P4}&\eqref{total power},\eqref{R_kk P2}-\eqref{fairness P2}.
\end{align}
\end{subequations}
To tackle this problem, we decompose (P4) into two subproblems. Since the active beamforming coefficients can still be optimized by solving (P3.2), we only need to focus on how to optimize the passive beamforming vector with the case of ${\Phi _2}$ or ${\Phi _3}$. In the following, we first invoke a novel SROCR approach~\cite{SROCR2017} to deal with the passive beamforming optimization with continuous phase shifts. Then, the passive beamforming design for discrete phase shifts is handled with the quantization-based scheme.
\vspace{-0.4cm}
\subsection{Passive Beamforming Optimization with ${\Phi _2}$}
Equipped with continuous phase shifters, each element on the IRS has a constant reflection amplitude, i.e. $\left| {{v_m}} \right|^2 = 1$. To tackle the unit modulus constraint, we define ${\mathbf{V}} = {\mathbf{v}}{{\mathbf{v}}^H}$ which satisfies ${{\mathbf{V}}} \succeq 0$, ${\rm {rank}}\left( {{{\mathbf{V}}}} \right) = 1$ and ${\left[ {\mathbf{V}} \right]_{mm}} = 1,m = 1,2, \cdots ,M + 1$. Then, for any given feasible active beamforming coefficients ${\left\{ {{\mathbf{w}}_k^l} \right\}}$, Problem (P4) can be written as
\vspace{-0.4cm}
\begin{subequations}
\begin{align}\label{P4.1}
({{\rm{P4.1}}}):&\mathop {\max }\limits_{{{\mathbf{V}}},\left\{ {{R_{k \to j}},{S_{kj}},{I_{kj}}} \right\}} \;\;\sum\limits_{k = 1}^K {{R_{k \to k}}}     \\
\label{S_jk P4.1}{\rm{s.t.}}\;\;&\frac{1}{{{S_{kj}}}} \le {\rm {Tr}}\left( {{\mathbf{V}}{{\mathbf{H}}_j}{{\mathbf{w}}_k}{\mathbf{w}}_k^H{\mathbf{H}}_j^H} \right),\forall k,j \in {{\mathcal{K}}},\\
\label{I_jk P4.1}&{I_{kj}} \ge \sum\limits_{\Omega \left( i \right) > \Omega \left( k \right)} {{\rm {Tr}}\left( {{\mathbf{V}}{{\mathbf{H}}_j}{{\mathbf{w}}_i}{\mathbf{w}}_i^H{\mathbf{H}}_j^H} \right)}  + {\sigma ^2},\forall k,j \in {{\mathcal{K}}},\\
\label{fairness P4.1}&
  {\rm {Tr}}\left( {{\mathbf{V}}{{\mathbf{H}}_j}{{\mathbf{w}}_k}{\mathbf{w}}_k^H{\mathbf{H}}_j^H} \right) \ge {\rm {Tr}}\left( {{\mathbf{V}}{{\mathbf{H}}_j}{{\mathbf{w}}_i}{\mathbf{w}}_i^H{\mathbf{H}}_j^H} \right), \Omega \left( k \right) < \Omega \left( i \right),\forall i,j,k, \\
\label{Vmm}&{\left[ {\mathbf{V}} \right]_{mm}} = 1,m = 1,2, \cdots ,M + 1,\\
\label{SDP V}&{{\mathbf{V}}}  \succeq  0, {\mathbf{V}} \in {{\mathbb{H}}^{M + 1}},\\
\label{rank 1 V}&{\rm {rank}}\left( {{{\mathbf{V}}}} \right) = 1,\\
\label{constraints P4.1}&\eqref{R_kk P2},\eqref{SIC P2}.
\end{align}
\end{subequations}
\vspace{-1cm}

\noindent The non-convexity of (P4.2) lies in the non-convex constraint \eqref{R_kk P2} and the rank-one constraint \eqref{rank 1 V}. Similarly, the non-convex constraint \eqref{R_kk P2} can be replaced with its lower bound in \eqref{Rjk lower bound}. For the rank-one constraint \eqref{rank 1 V}, the conventional approach is applying SDR~\cite{Wu2019IRS,Yang_ax}, where we first solve the problem by ignoring the rank-one constraint, and then construct a rank-one solution with Gaussian randomization method if the solution obtained from the relaxed problem is not rank-one. One drawback of this approach is that the constructed rank-one solution is normally a suboptimal solution or even infeasible for the original problem. The objective function's value may not be non-increasing after each iteration, which results in the convergence of the proposed algorithm cannot be guaranteed. Driven by this issue, we propose a novel SROCR-based algorithm to obtain a local optimal rank-one solution. The basic framework of the SROCR approach~\cite{SROCR2017} can be found in Appendix B. By replacing the non-convex term in \eqref{R_kk P2} with the lower bound in \eqref{Rjk lower bound}, Problem (P4.1) can be written as
\vspace{-0.4cm}
\begin{subequations}
\begin{align}\label{P4.2}
({{\rm{P4.2}}}):&\mathop {\max }\limits_{{{\mathbf{V}}},\left\{ {{R_{k \to j}},{S_{kj}},{I_{kj}}} \right\}} \;\;\sum\limits_{k = 1}^K {{R_{k \to k}}}     \\
\label{R_kk P4.2}{\rm{s.t.}}\;\;&{R_{k \to j}} \le R_{kj}^{low},\forall k,j \in {{\mathcal{K}}},\\
\label{rank 1 V P4.2}&{\rm {rank}}\left( {{{\mathbf{V}}}} \right) = 1,\\
\label{constraints P4.2}&\eqref{SIC P2},\eqref{S_jk P4.1}-\eqref{SDP V}.
\end{align}
\end{subequations}
\vspace{-1.2cm}

\noindent Now, Problem (P4.2) satisfies the general framework for the SROCR approach, which is shown in Problem (P) in Appendix B. Therefore, we apply the SROCR approach to solve Problem (P4.2). First, we replace the non-convex rank-one constraint with the relaxed convex constraint which controls the largest eigenvalue to trace ratio of ${\mathbf{V}}$ with the parameter ${\omega ^{\left( i \right)}} \in \left[ {0,1} \right]$. Thus, the relaxed optimization problem can be expressed as
\vspace{-0.4cm}
\begin{subequations}
\begin{align}\label{P4.3}
({{\rm{P4.3}}}):&\mathop {\max }\limits_{{{\mathbf{V}}},\left\{ {{R_{k \to j}},{S_{kj}},{I_{kj}}} \right\}} \;\;\sum\limits_{k = 1}^K {{R_{k \to k}}}     \\
\label{rank 1 V P4.3}{\rm{s.t.}}\;\;&{{\mathbf{u}}_{\max }}{\left( {{{\mathbf{V}}^{\left( i \right)}}} \right)^H}{\mathbf{V}}{{\mathbf{u}}_{\max }}\left( {{{\mathbf{V}}^{\left( i \right)}}} \right) \ge {\omega ^{\left( i \right)}}{\rm{Tr}}\left( {\mathbf{V}} \right),\\
\label{constraints P4.2}&\eqref{SIC P2},\eqref{S_jk P4.1}-\eqref{SDP V},\eqref{R_kk P4.2},
\end{align}
\end{subequations}
\vspace{-1.2cm}

\noindent where ${{\mathbf{u}}_{\max }}\left( {{{\mathbf{V}}^{\left( i \right)}}} \right)$ is the eigenvector corresponding to the largest eigenvalue of ${{{\mathbf{V}}^{\left( i \right)}}}$ and ${{{\mathbf{V}}^{\left( i \right)}}}$ is the obtained solution in the $i$th iteration with ${\omega ^{\left( i \right)}}$. It can be verified that Problem (P4.3) is a convex problem that can be efficiently solved with convex optimization software, such as CVX~\cite{cvx}. By increasing the parameter ${\omega ^{\left( i \right)}}$ from 0 to 1 after each iteration, we can derive a locally optimal rank-one solution for (P4.2). The details of solving Problem (P4.2) are summarized in \textbf{Algorithm 2}. Similarly, the objective function's value obtained from (P4.2) serves as a lower bound on that of (P4.1) due to the replacement of non-convex terms with their lower bounds. After solving (P4.2), the passive beamforming vector ${\mathbf{v}}$ for the continuous phase shifts case can be obtained through Cholesky decomposition, e.g. ${{\mathbf{V}}^{l + 1}} = {\mathbf{v}}{{\mathbf{v}}^H}$.
\begin{algorithm}[!h]\label{method1}
\caption{Iterative algorithm for solving Problem (P4.2)} 
 \hspace*{0.02in} 用来控制位置，同时利用 \\ 进行换行
\hspace*{0.02in}{Initialize convergence thresholds $\epsilon_1$, $\epsilon_2$ and feasible ${{{\mathbf{V}}^l}}$ to problem (P4.2), $i=0$.\\ Solve the relaxed problem (P4.3) with ${\omega ^{\left( i \right)}} = 0$, the obtained solution is denoted by ${{{\mathbf{V}}^{\left( i \right)}}}$, initialize the step size ${\delta ^{\left( i \right)}}$.}\\
\vspace{-0.4cm}
\begin{algorithmic}[1]
\STATE {\bf repeat}
\STATE Solve the convex Problem (P4.3) for given $\left\{ {{\omega ^{\left( i \right)}},{{\mathbf{V}}^{\left( i \right)}}} \right\}$.
\STATE {\bf if} Problem (P4.3) is solvable {\bf then}
\STATE The optimal solution is denoted by ${{{\mathbf{V}}^{\left( {i + 1} \right)}}}$.
\STATE ${\delta ^{\left( {i + 1} \right)}} = {\delta ^{\left( i \right)}}$.
\STATE {\bf else}
\STATE ${\delta ^{\left( {i + 1} \right)}} = {{{\delta ^{\left( i \right)}}} \mathord{\left/
 {\vphantom {{{\delta ^{\left( i \right)}}} 2}} \right.
 \kern-\nulldelimiterspace} 2}$.
\STATE {\bf end}
\STATE ${\omega ^{\left( {i + 1} \right)}} = \min \left( {1,\frac{{{\lambda _{\max }}\left( {{{\mathbf{V}}^{\left( {i + 1} \right)}}} \right)}}{{{\rm {Tr}}\left( {{{\mathbf{V}}^{\left( {i + 1} \right)}}} \right)}} + {\delta ^{\left( {i + 1} \right)}}} \right)$.
\STATE $i=i+1$.
\STATE {\bf until} $\left| {1 - {\omega ^{\left( {i - 1} \right)}}} \right| \le \epsilon_1$ and $\left| {{g_0}\left( {{{\mathbf{X}}^{\left( i \right)}}} \right) - {g_0}\left( {{{\mathbf{X}}^{\left( {i - 1} \right)}}} \right)} \right| \le \epsilon_2$, ${{{\mathbf{V}}^{l + 1}} = {{\mathbf{V}}^{\left( i \right)}}}$.
\end{algorithmic}
\end{algorithm}
\vspace{-0.4cm}
\subsection{Passive Beamforming Optimization with ${\Phi _3}$}
In this subsection, we discuss the passive beamforming design with discrete phase shifts. The optimization problem becomes a combinatorial optimization problem. Though the optimal solution can be obtained via an exhaustive search, it requires a prohibitive complexity since the number of elements on the IRS is usually large. Recall the fact that elements in ${\Phi _2}$ and ${\Phi _3}$ both follow the unit modulus constraint. Hence, we can directly quantize the obtained solution ${{\mathbf{v}}_{{\Phi _2}}}$ in the continuous phase shifts case to the nearest feasible point ${{\mathbf{v}}_{{\Phi _3}}}$ in the discrete phase shifts case as follows
\vspace{-0.4cm}
\begin{align}\label{quantize}
{v_{m,{\Phi _3}}} = \left\{ \begin{gathered}
  {e^{j\theta _m^*}},\;\;\;\;m = 1, \cdots ,M, \hfill \\
  1,\;\;\;\;\;\;\;\;m = M + 1, \hfill \\
\end{gathered}  \right.\;
\end{align}
\vspace{-0.8cm}

\noindent where $\theta _m^* = \arg \;\;\mathop {\min }\limits_{\theta  \in {\mathcal{D}}} \left| {\theta  - {\rm {angle}}\left( {{v_{m,{\Phi _2}}}} \right)} \right|.$ However, the obtained solution ${{\mathbf{v}}_{{\Phi _3}}}$ with quantization method may be not a locally optimal solution. In order to make the objective function's value to be non-increasing after each iteration for ${\Phi _3}$, we update ${{\mathbf{v}}_{{\Phi _3}}}$ only when ${\eta _{\Phi 3}}\left( {\left\{ {{\mathbf{w}}_k^{l + 1}} \right\},{\mathbf{v}}_{{\Phi _3}}^{l + 1}} \right) \ge {\eta _{\Phi 3}}\left( {\left\{ {{\mathbf{w}}_k^{l + 1}} \right\},{\mathbf{v}}_{{\Phi _3}}^l} \right)$.
\vspace{-0.4cm}
\subsection{Proposed Algorithm, Complexity and Convergence}
Similarly, we propose an iterative algorithm for Problem (P4) with non-ideal IRS by utilizing the AO method. The details of the proposed algorithm are summarized in \textbf{Algorithm 3}. The complexity of the SDP subproblem for the passive beamforming optimization is ${{\mathcal{O}}}\left( {I_{ite}^S\left( {\max {{\left( {M + 1,3K\left( {K - 1} \right)} \right)}^4}\sqrt {M + 1} \log \frac{1}{\varepsilon }} \right)} \right)$, where $I_{ite}^S$ denotes the number of iterations for Algorithm 2 with the SROCR approach and $\varepsilon$ denotes the accuracy. The total complexity of Algorithm 3 is
\[{{\mathcal{O}}}\left(\! {I_{ite}^N\left(\! {\left(\! {\max {{\left( {N,3K\!\left( {K\! -\! 1} \right)} \right)}^4}\sqrt N \log \frac{1}{\varepsilon }} \right) \!+\! I_{ite}^S\left(\! {\max {{\left( {M\! +\! 1,3K\!\left( {K\! - \!1} \right)} \right)}^4}\sqrt {M \!+\! 1} \log \frac{1}{\varepsilon }} \right)}\! \right)}\! \right),\]
where $I_{ite}^N$ denotes the number of iterations for Algorithm 3\textcolor{black}{~\cite{Luo}}. It can be seen that the computational complexity of the algorithm with the non-ideal IRS is larger than that with the ideal IRS. The convergence of Algorithm 3 can be proved in a similar way as Algorithm 1.
\begin{algorithm}[!t]\label{method3}
\caption{Proposed SROCR-based algorithm for solving Problem (P4) with $\Phi_2$ or $\Phi_3$} 
 \hspace*{0.02in} 用来控制位置，同时利用 \\ 进行换行
\hspace*{0.02in} {Initialize a decoding order $\Omega $ and feasible solutions $\left\{ {{\mathbf{w}}_k^l} \right\},{{\mathbf{v}}^l}$ to (P4), $l=0$.}\\
\vspace{-0.4cm}
\begin{algorithmic}[1]
\STATE {\bf repeat}
\STATE Solve Problem (P3.2) for given ${\mathbf{v}}^l$, where the optimal solution is denoted by $\left\{ {{\mathbf{w}}_k^{l + 1}} \right\}$.
\STATE Solve Problem (P4.2) with the proposed \textbf{Algorithm 2} for given $\left\{ {{\mathbf{w}}_k^{l + 1}} \right\}$, where the optimal solution is denoted by ${\mathbf{v}}^{l+1}$.
\STATE {\bf if} $\Phi  = {\Phi _3}$ {\bf then}
\STATE Construct the feasible solution ${{\mathbf{v}}_{{\Phi _3}}}$ via \eqref{quantize} with ${\mathbf{v}}^{l+1}$.
\STATE $\;\;\;\;$    {\bf if} ${\eta _{\Phi 3}}\left( {\left\{ {{\mathbf{w}}_k^{l + 1}} \right\},{\mathbf{v}}_{{\Phi _3}}^{l + 1}} \right) \ge {\eta _{\Phi 3}}\left( {\left\{ {{\mathbf{w}}_k^{l + 1}} \right\},{\mathbf{v}}_{{\Phi _3}}^l} \right)$ {\bf then}
\STATE $\;\;\;\;$       ${{\mathbf{v}}^{l + 1}} = {{\mathbf{v}}_{{\Phi _3}}}$.
\STATE $\;\;\;\;$    {\bf else}
\STATE $\;\;\;\;$       ${{\mathbf{v}}^{l + 1}} = {{\mathbf{v}}^l}$.
\STATE $\;\;\;\;$    {\bf end}
\STATE {\bf end}
\STATE $l=l+1$.
\STATE {\bf until} the fractional increase of the objective value is below a threshold $\xi   > 0$.
\end{algorithmic}
\end{algorithm}
\begin{remark}\label{Upper bound}
\textcolor{black}{\emph{In practical applications, it is costly to deploy IRSs with ideal case or continuous phase shifts due to hardware limitations. However, it is still important to analyze the cases of $\Phi_1$ and \textcolor{black}{$\Phi_ 2$} since they provide theoretical performance upper bounds to the case of \textcolor{black}{$\Phi_ 3$}.}}
\end{remark}
\vspace{-0.6cm}
\subsection{Passive Beamforming Optimization with 1-Bit Phase Shifters}
As described earlier, the quantization method can provide a feasible solution for the passive beamforming design in the discrete phase shifts case. However, this method may experience substantial performance losses with low resolution phase shifters, e.g., $B=1$. To overcome this drawback, we investigate the passive beamforming design with 1-bit phase shifters in this subsection. Specifically, the passive beamforming optimization problem of 1-bit phase shifters under given ${\left\{ {{\mathbf{w}}_k} \right\}}$ can be written as
\vspace{-0.4cm}
\begin{subequations}
\begin{align}\label{P4.4}
({{\rm{P4.4}}}):&\mathop {\max }\limits_{{\mathbf{v}},\left\{ {{R_{k \to j}},{S_{kj}},{I_{kj}}} \right\}} \;\;\sum\limits_{k = 1}^K {{R_{k \to k}}}     \\
\label{discrete phase shift P4.4}{\rm{s.t.}}\;\;&{v_m} \in \left\{ { - 1,1} \right\},m = 1,2, \cdots ,M,{v_{M + 1}} = 1,\\
\label{constraints P4.4}&\eqref{R_kk P2}-\eqref{fairness P2}.
\end{align}
\end{subequations}
\vspace{-1.2cm}

\noindent Problem (P4.4) is a Boolean quadratic problem (BQP). Similarly, let ${\mathbf{V}} = {\mathbf{v}}{{\mathbf{v}}^H}$ which satisfies ${\mathbf{V}} \in {{\mathbb{S}}^{M + 1}}$, ${{\mathbf{V}}} \succeq 0$, ${\rm {rank}}\left( {{{\mathbf{V}}}} \right) = 1$ and ${\left[ {\mathbf{V}} \right]_{mm}} = 1,m = 1,2, \cdots ,M + 1$. Then, Problem (P4.4) can be expressed as
\vspace{-0.4cm}
\begin{subequations}
\begin{align}\label{P4.5}
({{\rm{P4.5}}}):&\mathop {\max }\limits_{{\mathbf{V}},\left\{ {{R_{k \to j}},{S_{kj}},{I_{kj}}} \right\}} \;\;\sum\limits_{k = 1}^K {{R_{k \to k}}}     \\
\label{R_kk P4.5}{\rm{s.t.}}\;\;&{R_{k \to j}} \le R_{kj}^{low},\forall k,j \in {{\mathcal{K}}},\\
\label{SDP V P4.5}&{{\mathbf{V}}}  \succeq  0, {\mathbf{V}} \in {{\mathbb{S}}^{M + 1}},\\
\label{rank 1 V P4.5}&{\rm {rank}}\left( {{{\mathbf{V}}}} \right) = 1,\\
\label{constraints P4.4}&\eqref{SIC P2},\eqref{S_jk P4.1}-\eqref{Vmm}.
\end{align}
\end{subequations}
\vspace{-1.2cm}

\noindent Problem (P4.5) can be regarded as a non-convex rank-one optimization problem with a real symmetric matrix set, which can also be solved by the proposed SROCR-based algorithm.
\begin{figure}[b!]
    \begin{center}
        \includegraphics[width=3in]{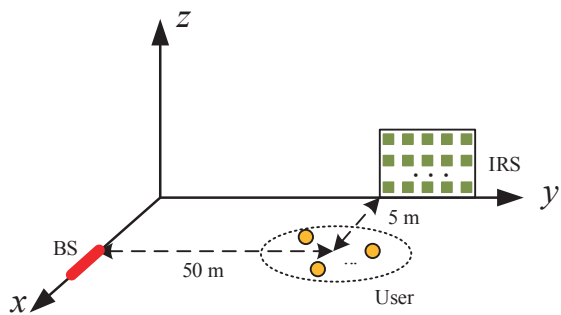}
        \caption{\textcolor{black}{Simulation setup of the MISO IRS-aided NOMA system.}}
        \label{simulation setup}
    \end{center}
\end{figure}
\begin{remark}\label{Upper bound 1bit}
\textcolor{black}{\emph{Since a locally optimal rank-one solution can be obtained for the passive beamforming optimization with the SROCR approach, the sum rate of 1-bit phase shifters obtained by the SROCR-based algorithm will be no worse than that of the quantization-based scheme. }}
\end{remark}
\vspace{-0.6cm}
\section{Simulation Results}
In this section, simulation results are provided to validate the effectiveness of the proposed algorithms. \textcolor{black}{As illustrated in Fig. \ref{simulation setup}, we consider a three-dimensional (3D) coordinate system, where the BS and IRS are equipped with a uniform linear array (ULA) located on the $x$-axis and a uniform planar array (UPA) located parallel to the $y-z$ plane, respectively. The reference antennas at the BS and IRS are set at $\left( {5,0,0} \right)$ meters and $\left( {0,50,0} \right)$ meters, respectively. The antenna spacing is half wavelength. The number of IRS elements is set as $M = {M_y}{M_z}$, where ${M_y}$ and ${M_z}$ denote the number of IRS elements along the $y$-axis and $z$-axis, respectively. We set ${M_y}=5$ and increase ${M_z}$ linearly with $M$. The served users $k \in {\mathcal{K}}$ are randomly and uniformly distributed in a circle region  centered at $\left( {50,5,0} \right)$ with the radius of 3 m. The distances for the direct BS-user link, the BS-IRS link and the IRS-user link are denoted by ${d_{BU,k}}$, ${d_{BI}}$ and ${d_{IU,k}}$, respectively. The distance-dependent path loss for all channels is modeled as}
\vspace{-0.4cm}
\begin{align}\label{path loss}
\textcolor{black}{PL\left( d \right) = {\rho _0}{\left( {\frac{d}{{{d_0}}}} \right)^{ - \alpha }},}
\end{align}
\vspace{-1.0cm}

\noindent \textcolor{black}{where ${\rho _0} =  - 30$ dB denotes the path loss at the reference distance ${d_0} = 1$ meter, $d$ denotes the link distance and $\alpha$ denotes the path loss exponent. For small scale fading, the Rayleigh fading channel model and the Rician fading model are assumed for the direct BS-user link and the BS-IRS/IRS-user links, respectively. Then, the corresponding channel coefficients can be expressed as}
\vspace{-0.4cm}
\begin{subequations}
\begin{align}\label{channel coefficients1}
&\textcolor{black}{{{\mathbf{h}}_k} = \sqrt {PL\left( {{d_{BU,k}}} \right)} {\mathbf{h}}_k^{{\rm{NLoS}}},k \in {\mathcal{K}},}\\
\label{channel coefficients2}&\textcolor{black}{{\mathbf{G}} = \sqrt {\frac{{PL\left( {{d_{BI}}} \right)}}{{{K_{BI}} + 1}}} \left( {\sqrt {{K_{BI}}} {\mathbf{G}}_{}^{{\rm{LoS}}} + {\mathbf{G}}_{}^{{\rm{NLoS}}}} \right),}\\
\label{channel coefficients3}&\textcolor{black}{{{\mathbf{r}}_k} = \sqrt {\frac{{PL\left( {{d_{IU,k}}} \right)}}{{{K_{IU}} + 1}}} \left( {\sqrt {{K_{IU}}} {\mathbf{r}}_k^{{\rm{LoS}}} + {\mathbf{r}}_k^{{\rm{NLoS}}}} \right),k \in {\mathcal{K}},}
\end{align}
\end{subequations}
\vspace{-1.0cm}

\noindent \textcolor{black}{where ${{K_{BI}}}$ and ${{K_{IU}}}$ denote the Rician factors, ${{\mathbf{G}}^{{\rm{LoS}}}}$ and ${{\mathbf{r}}_k^{\rm{LoS}}}$ denote the deterministic LoS components, ${\mathbf{h}}_k^{{\rm{NLoS}}}$, ${{\mathbf{v}}_{}^{{\rm{NLoS}}}}$ and ${{\mathbf{g}}_k^{{\rm{NLoS}}}}$ denote the Rayleigh fading components. From \eqref{channel coefficients2} and \eqref{channel coefficients3}, it can be observed that the reflection link suffers from the ``double-fading'' effect~\cite{Griffin}. In this paper, the path loss exponents for the direct link, AP-IRS link and IRS-user link are set to be ${\alpha _D} = 3.5$, ${\alpha _{BI}} = 2.2$ and ${\alpha _{IU}} = 2.2$, respectively. The Rician factors are ${K_{BI}} = {K_{IU}} = 3$ dB. The noise power is given by ${\sigma ^2} = B{N_0}$ with bandwidth $B$ = 1 MHz and the effective noise power density of $N_0 = -150$ dbm/Hz~\cite{Wu2019IRS}.} All simulation results are obtained by averaging over 100 independent channel realizations. \textcolor{black}{To strike a balance between the obtained performance and computational complexity, the convergence threshold of the proposed algorithms is set to be $\xi  = {10^{ - 2}}$.}
\vspace{-0.6cm}
\subsection{Convergence of Proposed Algorithms}
\begin{figure}[t!]
    \begin{center}
        \includegraphics[width=3in]{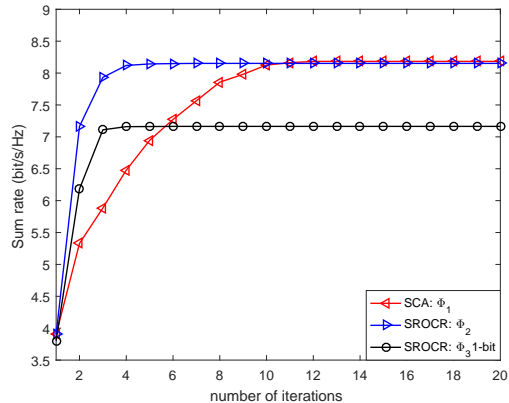}
        \caption{\textcolor{black}{Sum rate versus the number of iterations for $N=2$, $M=30$, $K=4$ and $P_T =  10$dBm.}}
        \label{Convergence}
    \end{center}
\end{figure}

\indent \textcolor{black}{In Fig. \ref{Convergence}, we first provide the convergence of Algorithm 1 and Algorithm 3 for $N=2$, $M=30$, $K=4$ and $P_T =  10$dBm.} The initial active beamforming coefficients $\left\{ {{\mathbf{w}}_k^0} \right\}$ and passive beamforming vector ${{\mathbf{v}}^0}$ are obtained with the following method\footnote{\textcolor{black}{It is worth noting that more sophisticated initialization schemes may further enhance the convergence speed and the attainable performance of the proposed algorithms, but this is beyond the scope of this paper.}}
\begin{itemize}
  \item \textbf{Passive beamforming initialization}: In the $\Phi_1$ and $\Phi_2$ cases, the phase shift of each element is uniformly distributed between $\left[ {0,2\pi } \right)$ and the reflection amplitude is set to be 1. In the $\Phi_3$ case, the initial passive beamforming vector is obtained by quantizing the random continuous phase shifts to the discrete phase shifts.
  \item \textbf{Active beamforming initialization}: Given the initial passive beamforming vector, the active beamforming vectors are generated with equal power allocation $\frac{{{P_T}}}{K}$, while satisfying the SIC decoding rate conditions and user rate fairness constraints.
\end{itemize}
\textcolor{black}{From Fig. \ref{Convergence}, it is observed that the sum rate of the proposed algorithms increase quickly with the number of iterations. The proposed SCA and SROCR based algorithms converge with around 6 and 12 iterations, respectively, which is consistent with \textbf{Remark \ref{convergence}}.}
\vspace{-0.6cm}
\subsection{Impact of the IRS}
In order to demonstrate the benefits brought by deploying the IRS, we compare the proposed algorithms with the following benchmark schemes
\begin{itemize}
  \item \textcolor{black}{\textbf{SDR approach}: In this case, the passive beamforming vector in $\Phi_2$ is optimized by invoking the SDR approach which ignores the non-convex rank-one constraint~\cite{Wu2019IRS,Yang_ax}. The Gaussian randomization approach should be applied if the obtained solution is not rank-one.}
  \item \textbf{Random phase shifts}: In this case, the phase shifts of IRS elements are set with random values in $\Phi_2$. Then, we only optimize the active beamforming at the BS with the combined channels by solving Problem (P3.2).
  \item \textbf{Without IRS}: In this case, the BS served multiple users without the aid of an IRS. The active beamforming vectors are optimized with BS-user channels by solving Problem (P3.2).
\end{itemize}

\begin{figure}[b!]
\centering
\begin{minipage}[t]{0.45\linewidth}
\includegraphics[width=3in]{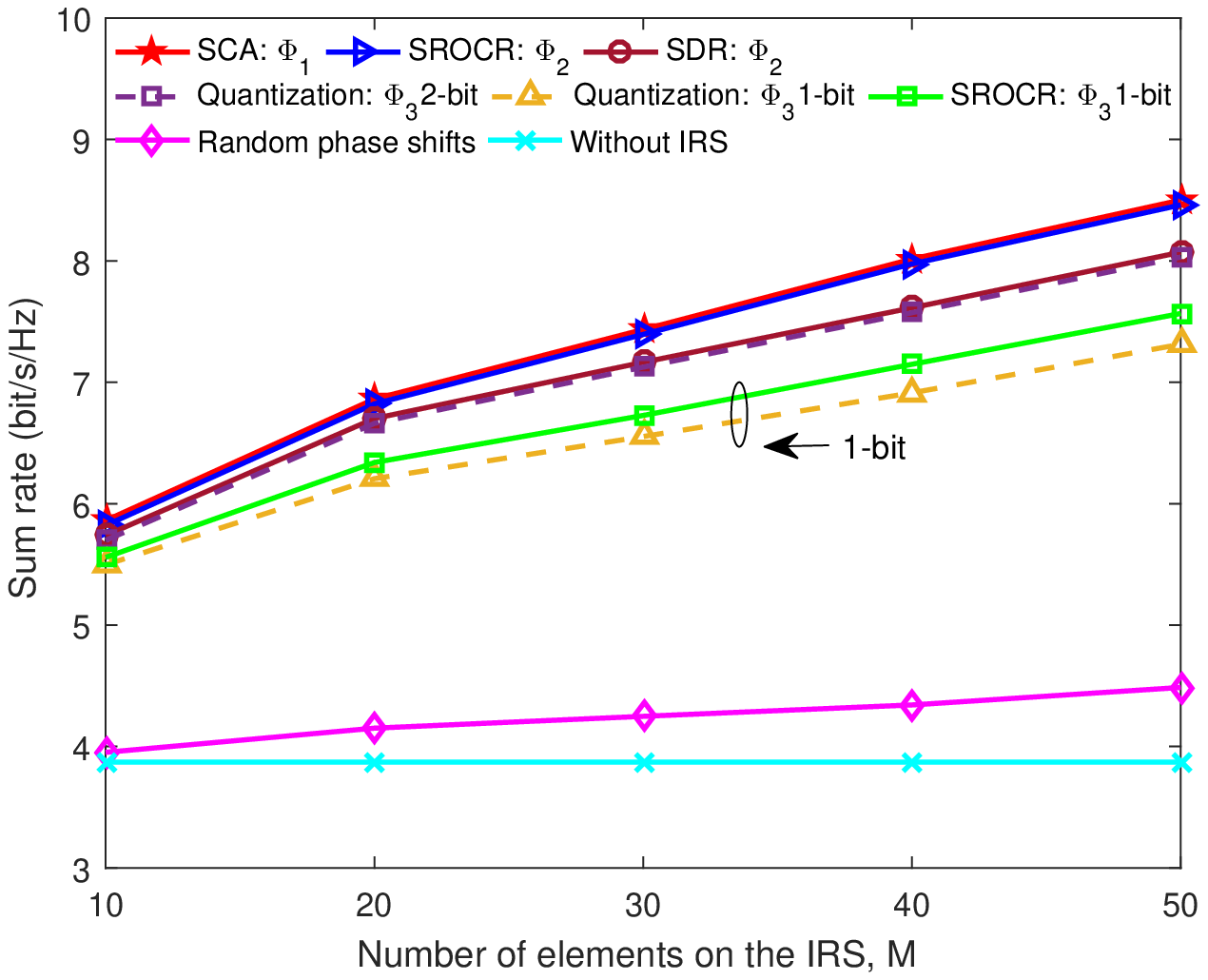}
\caption{\textcolor{black}{Sum rate versus the number of IRS elements $M$ for $N=2$, $K=4$, $P_T=10$dBm.}}
\label{element number}
\end{minipage}
\quad
\begin{minipage}[t]{0.45\linewidth}
\includegraphics[width=3in]{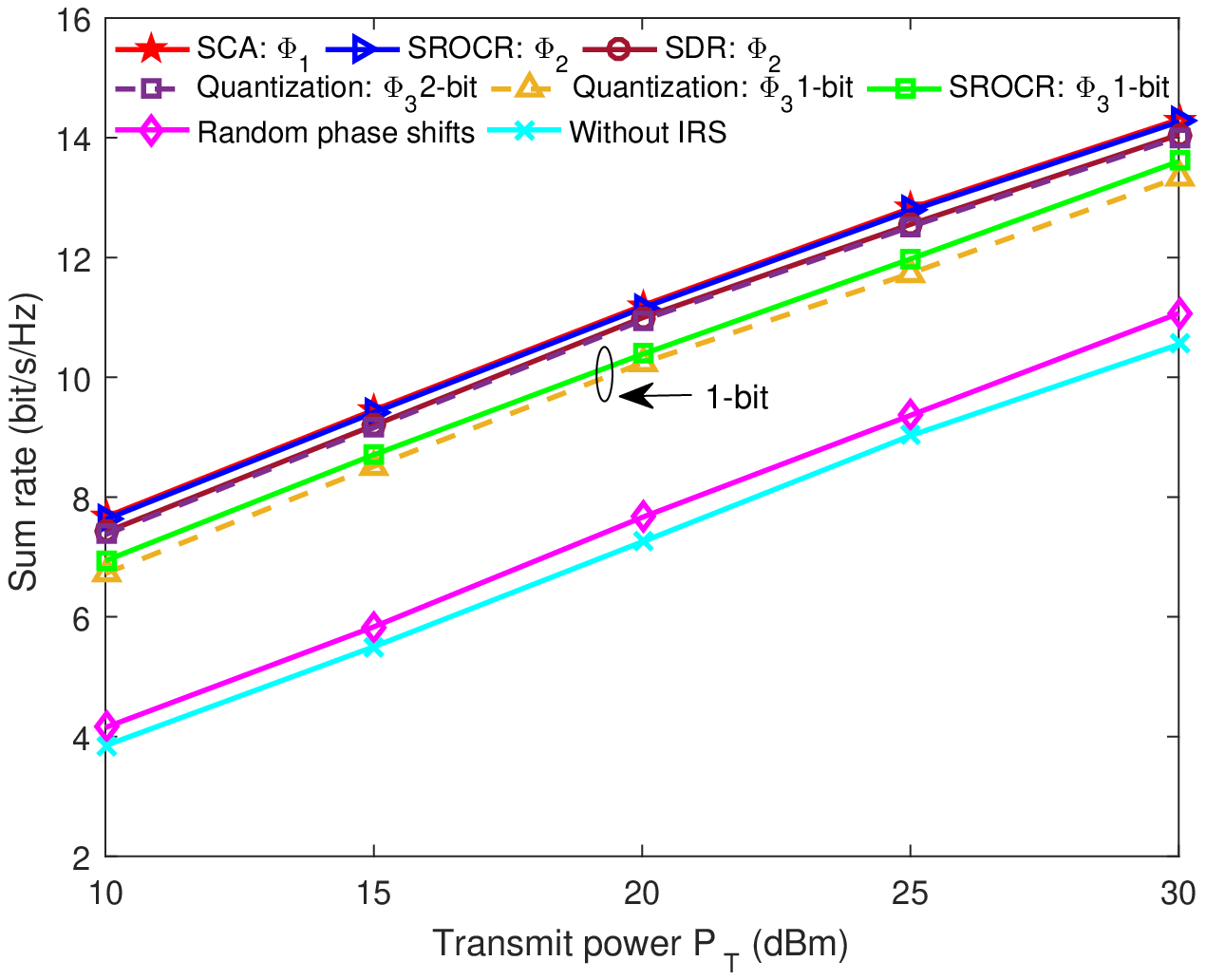}
\caption{\textcolor{black}{Sum rate versus transmit power $P_T$ for $N=2$, $M=30$, $K=4$.}}
\label{transmit power}
\end{minipage}
\end{figure}
\subsubsection{Sum Rate versus the Number of IRS Reflecting Elements} In Fig. \ref{element number}, we provide the sum rate versus the number of IRS reflecting elements $M$ for $N=2$, $K=4$ and $P_T =  10$ dBm with different schemes. It is firstly observed that the sum rate of all IRS-aided schemes increase with the increase of $M$, while the sum rate of the ``Without IRS'' scheme remains unchanged. This is expected since a higher array gain can be achieved with a larger number of IRS reflecting elements. \textcolor{black}{Moreover, the proposed schemes and the ``SDR approach'' both significantly outperform the ``Random phase shifts'' scheme, which unveils the importance of IRS phase shifts optimization. It is also observed that the performance gain of the proposed SROCR algorithm over the ``SDR approach'' becomes more pronounced as the number of reflecting elements increases. This is expected since the probability of the obtained solution satisfying the rank-one constraints becomes lower when $M$ increases. The Gaussian randomization approach only provides an approximate solution, which results in performance degradation.} From the perspective of different IRS assumptions, it is observed that the performance gaps between the case of ideal IRS $\Phi_1$ and continuous phase shifts $\Phi _2$ can be ignored. This is because the amplitudes of the passive beamforming vectors obtained from Algorithm 1 are nearly one. For practical discrete phase shifts $\Phi_3$, we observe that the performance degradation caused by finite resolution phase shifters decreases as the resolution bit increases. In particular, the performance achieved by the SROCR-based algorithm with 1-bit phase shifters outperforms the quantization-based scheme, \textcolor{black}{which is consistent with \textbf{Lemma \ref{Upper bound 1bit}}.} This is because the SROCR-based algorithm is capable of finding a locally optimal rank-one solution.
\subsubsection{Sum Rate versus Transmit Power} Fig. \ref{transmit power} presents the achieved system sum rate versus the transmit power of different schemes for $N=2$, $M = 30$ and $K=4$. As illustrated, it is firstly observed that the sum rate of all considered schemes increase with the increase of $P_T$. However, for achieving the same sum rate, the proposed schemes consume less transmit power than the ``Without IRS'' scheme. \textcolor{black}{The ``SDR approach'' only achieves a similar performance as the 2-bit phase shifts case, which also verifies the effectiveness of the proposed SROCR-based algorithm.} It is also observed that the ``Random phase shifts'' scheme only slightly outperforms the ``Without IRS'' scheme since the IRS phase shifts are not optimized.
\begin{figure}[b!]
\centering
\begin{minipage}[t]{0.45\linewidth}
\includegraphics[width=3in]{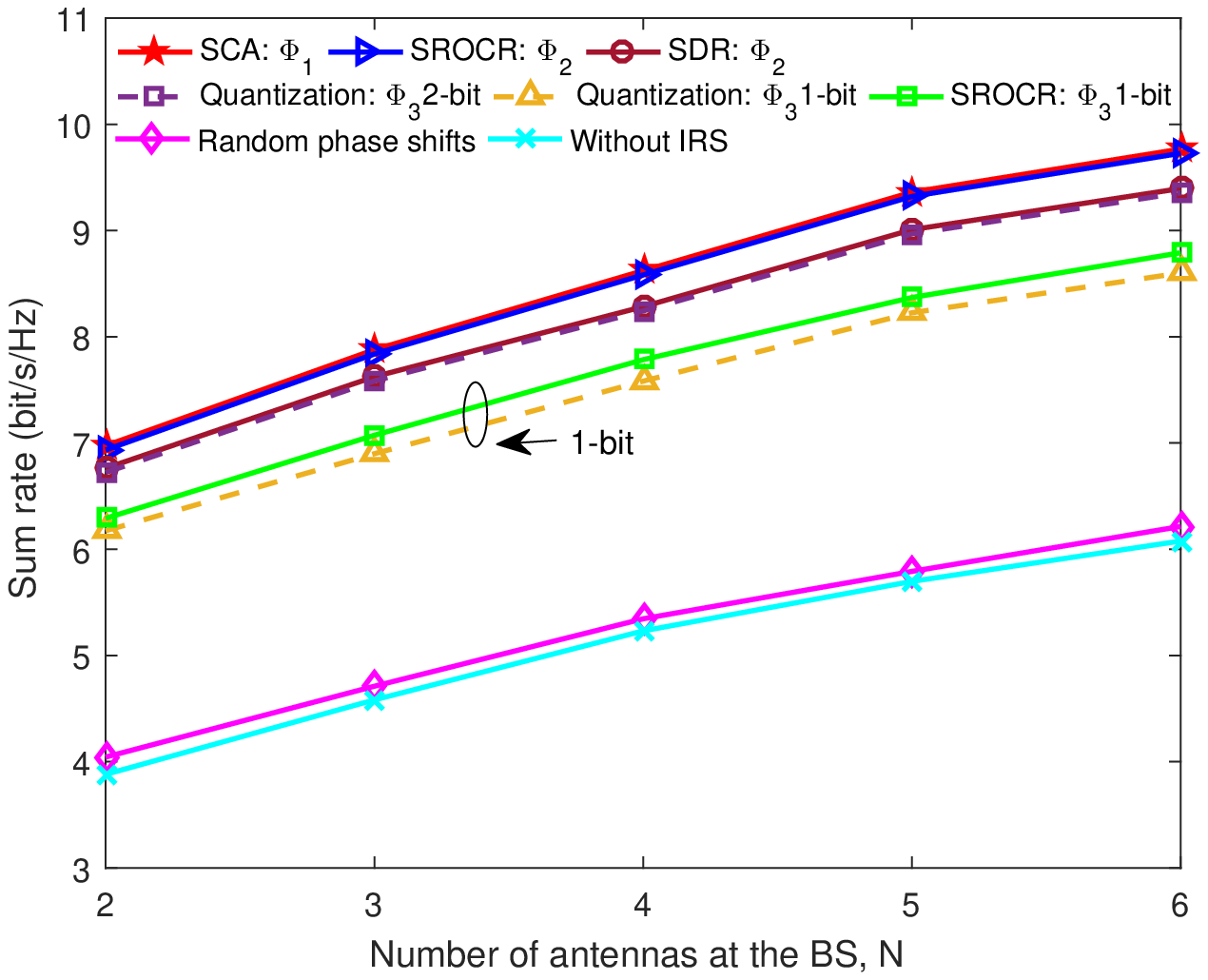}
\caption{\textcolor{black}{Sum rate versus the number of BS antennas $N$ for $M=30$, $K=4$, $P_T=10$dBm.}}
\label{antennas number}
\end{minipage}
\quad
\begin{minipage}[t]{0.45\linewidth}
\includegraphics[width=3in]{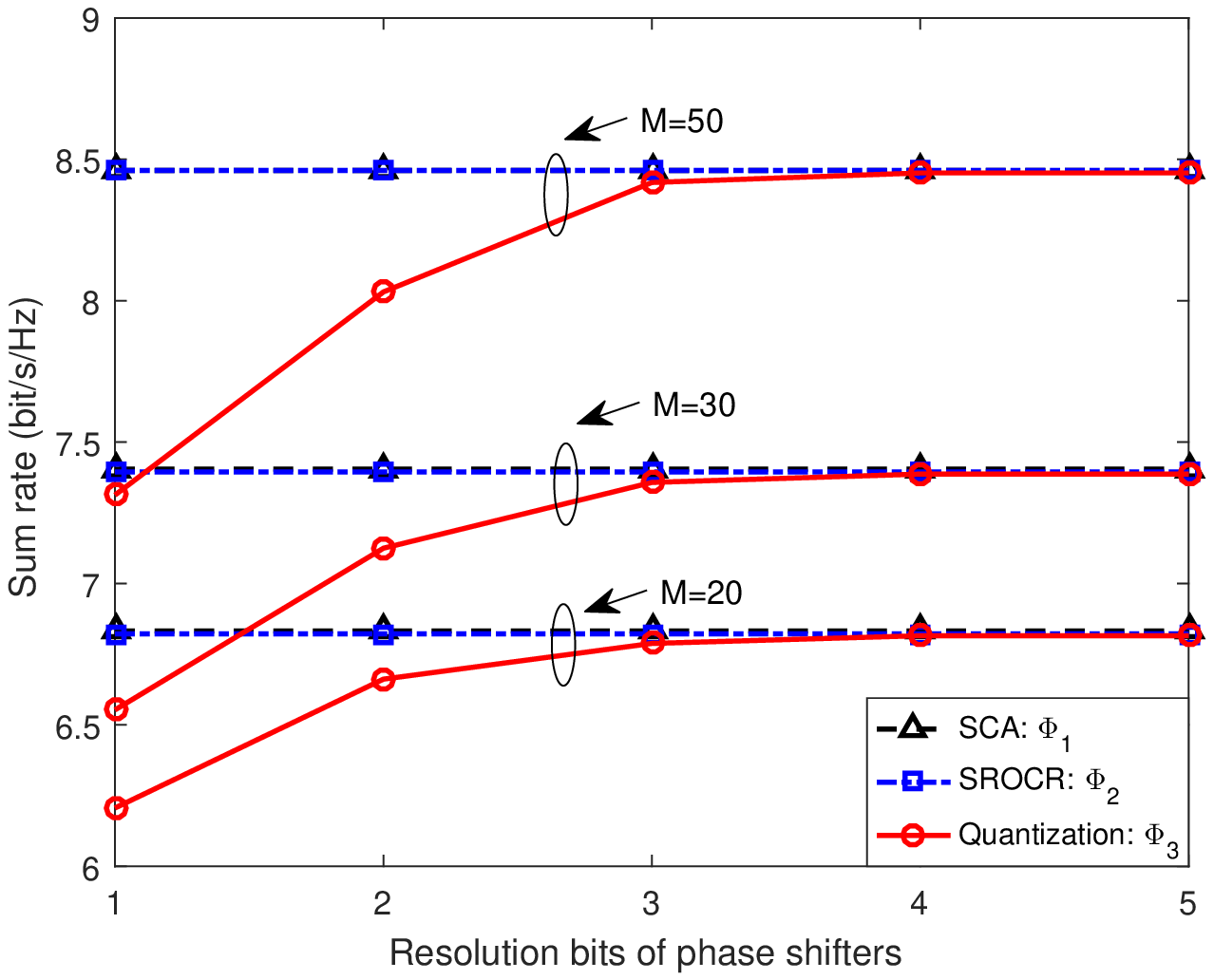}
\caption{\textcolor{black}{Sum rate versus resolution bits for $N=2$, $K=4$, $P_T=10$dBm.}}
\label{quan}
\end{minipage}
\end{figure}
\textcolor{black}{\subsubsection{Sum Rate versus the Number of BS Antennas} Fig. \ref{antennas number} provides the achieved system sum rate versus the number of BS antennas for $M = 30$, $K=4$ and $P_T =  10$ dBm with different schemes. The sum rate of all schemes increase with the increase of $N$ since a larger active beamforming gain can be achieved. It is also observed that the proposed algorithms significantly outperform the ``Random phase shifts'' and ``Without IRS'' schemes, which underscores the importance of joint active and passive beamforming optimization.}
\subsubsection{Sum Rate versus the Resolution Bits} Fig. \ref{quan} illustrates the sum rate versus the number of resolution bits of the IRS phase shifters. The ideal IRS case achieves the best performance, while the discrete case of $\Phi_3$ achieves the worst performance. This is expected since ${\Phi _1} \supseteq {\Phi _2} \supseteq {\Phi _3}$, \textcolor{black}{which is also consistent with \textbf{Lemma \ref{Upper bound}}.} It is also observed that the performance gap between $\Phi_3$ and $\Phi_1$ becomes narrower with the increase of resolution bits. As illustrated, ``1-bit'' and ``2-bit'' schemes are capable of achieving around 90\% and 95\% performance of the ideal IRS case, respectively. Moreover, the performance loss between the ``3-bit'' scheme and the ideal IRS is negligible.
\begin{figure}[b!]
\centering
\begin{minipage}[t]{0.45\linewidth}
\includegraphics[width=3in]{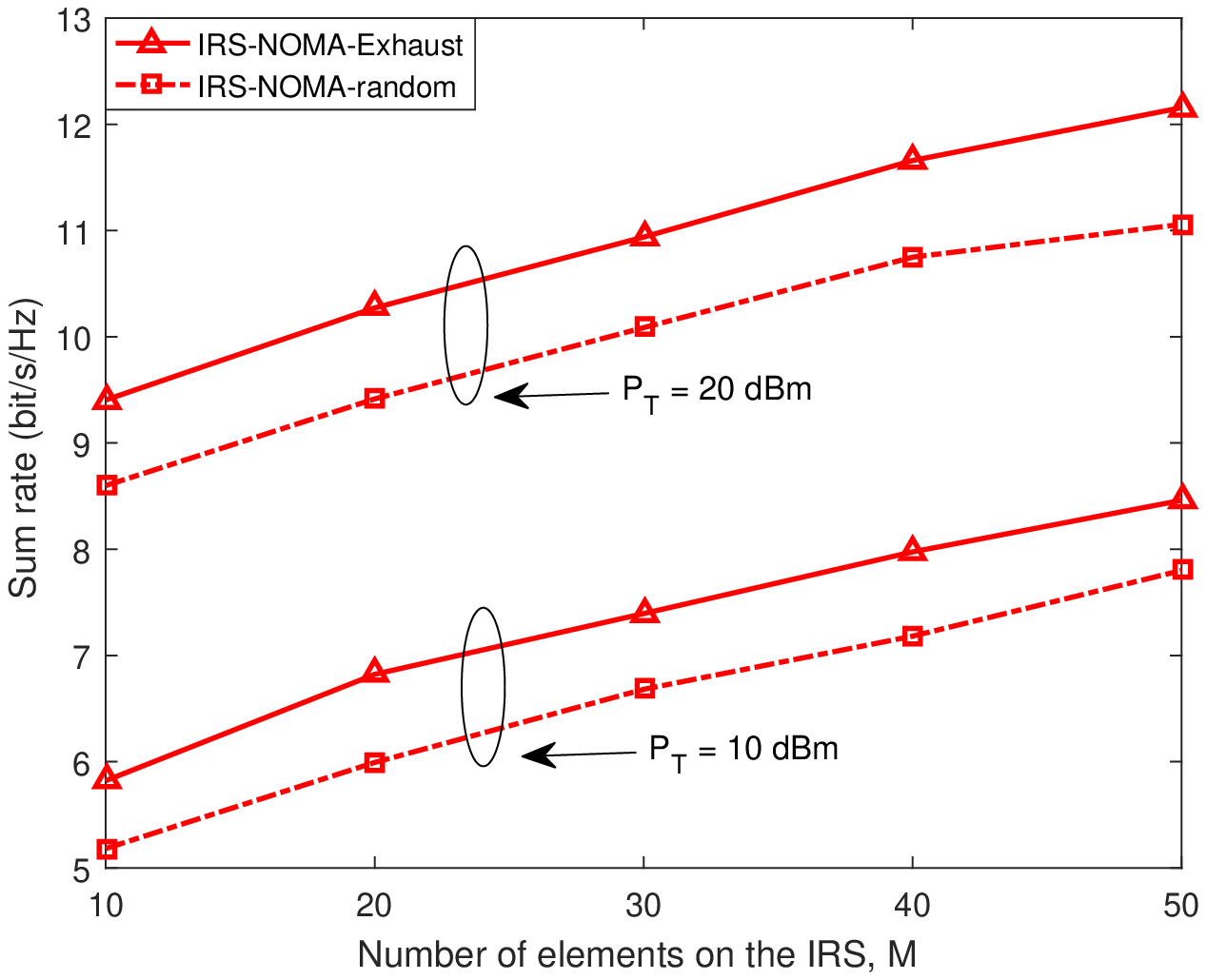}
\caption{\textcolor{black}{Sum rate versus the number of IRS elements $M$ with different decoding order schemes for $N=2$, $K=4$.}}
\label{random}
\end{minipage}
\quad
\begin{minipage}[t]{0.45\linewidth}
\includegraphics[width=3in]{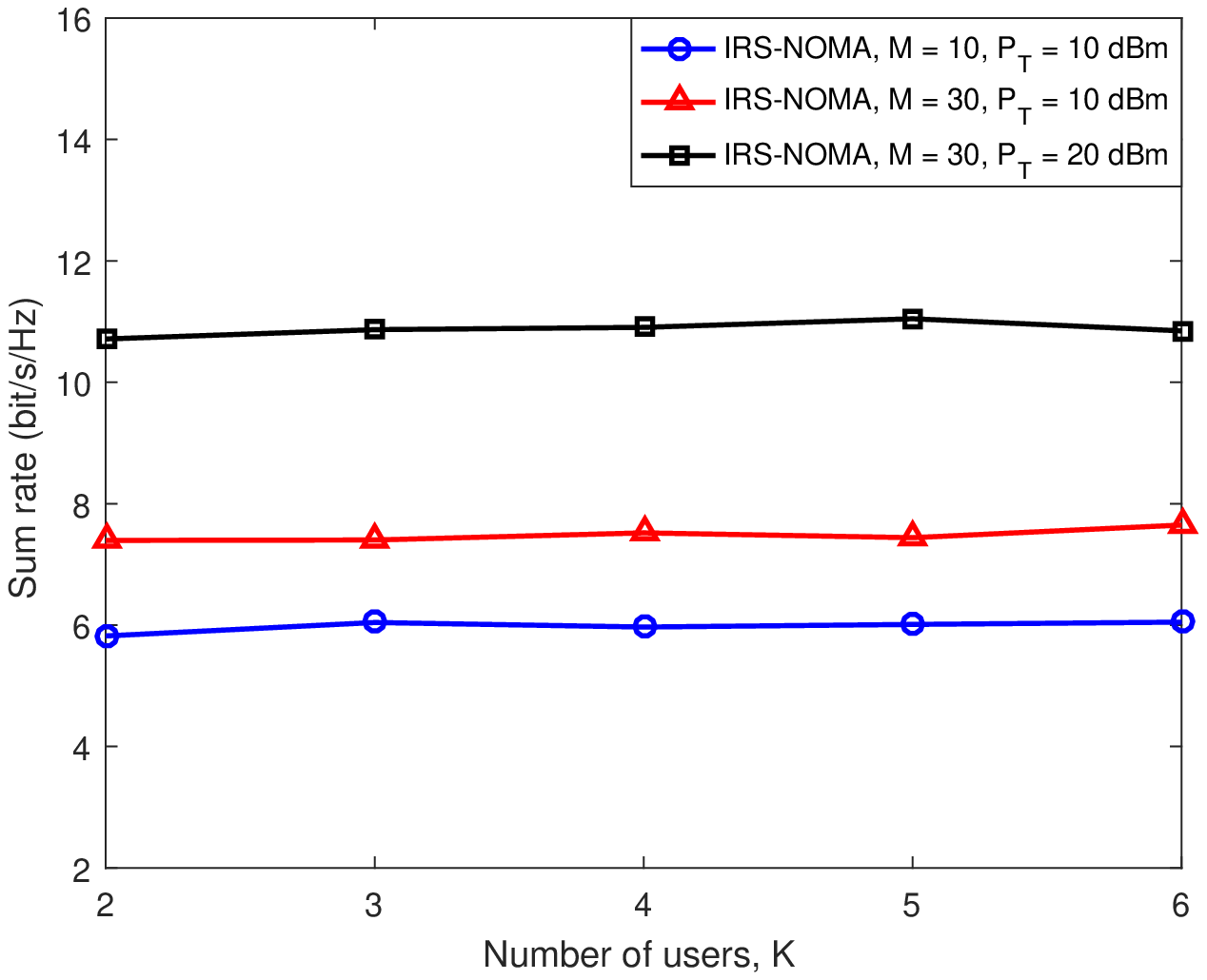}
\caption{\textcolor{black}{Sum rate versus the number of users $K$ for $N=2$.}}
\label{user number}
\end{minipage}
\end{figure}
\vspace{-0.4cm}
\subsection{Impact of Decoding Order}
To evaluate the impact of the decoding order on the sum rate, we compare the following schemes: 1) ``IRS-NOMA-Exhaust'' denotes the proposed algorithms where the optimal decoding order is selected through exhaustive search; 2) ``IRS-NOMA-Random'' denotes the case where the active and passive beamforming vectors are optimized with a randomly selected decoding order. For both schemes, continuous phase shifts are assumed and $N=2$, $K=4$. As illustrated in Fig. \ref{random}, the proposed algorithm outperforms the benchmark scheme, which highlights the importance of finding the optimal decoding order. It is worth noting that the proposed algorithms need to search over $K!$ possible decoding orders, which is acceptable when the number of users is not large. However, when $K$ is large, the complexity will be prohibitively high. In this case, some low-complexity ordering methods are required.
\vspace{-0.6cm}
\textcolor{black}{\subsection{Impact of the Number of Users}}
\textcolor{black}{In Fig. \ref{user number}, we investigated the sum rate versus the number of users for $N=2$, where continuous phase shifts are considered. It is observed that the obtained sum rate remains at an almost constant value with the increase of $K$. Though the impact of the number of users on the sum rate is negligible, it implies that a larger number of users results in a lower average user rate. Moreover, it is also found that a higher transmit power or larger number of IRS elements leads to a higher sum rate.}
\vspace{-0.6cm}
\subsection{Performance Comparison with OMA}
\begin{figure}[htb]
    \begin{center}
        \includegraphics[width=3in]{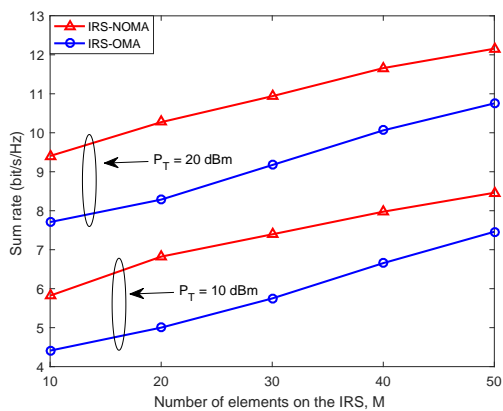}
        \caption{\textcolor{black}{Sum rate versus the number of IRS elements $M$ with different multiple access schemes for $N=2$, $K=4$.}}
        \label{OMA}
    \end{center}
\end{figure}

Finally, we compare the proposed IRS-aided NOMA scheme with the IRS-aided OMA scheme. In the IRS-aided OMA scheme, the BS serves all users through time division multiple access with the aid of the IRS, where the active and passive beamforming vectors are jointly optimized to maximize the received signal strength at the served user during each time slot. Fig. \ref{OMA} depicts the sum rate of both IRS-aided NOMA and IRS-aided OMA schemes for $N=2$ and $K=4$ in the continuous phase shifts case. It is observed that the IRS-aided NOMA scheme significantly outperforms the IRS-aided OMA scheme. This is expected since all users can be served simultaneously through the NOMA protocol compared with the OMA scheme.
\vspace{-0.6cm}
\vspace{-1.2cm}
\section{Conclusion}
In this paper, the MISO IRS-aided NOMA system has been investigated. The sum rate maximization problem was formulated by jointly optimizing the active and passive beamforming vectors under various IRS elements assumptions. To tackle these non-convex problems, the original problem was decoupled into two subproblems, namely, active beamforming optimization and passive beamforming optimization. For the ideal IRS case, the two non-convex subproblems were alternately solved by applying the SCA technique. For the non-ideal IRS case, a novel SROCR-based algorithm was proposed to find a locally optimal rank-one solution for continuous phase shifts. A quantization-based scheme was designed for discrete phase shifts. Simulation results showed that significant performance gains can be achieved by the proposed designs compared with non-IRS or IRS-aided OMA systems. \textcolor{black}{It is worth noting that obtaining perfect CSI is a challenging task for IRS-aided networks, our future research will consider robust beamforming design with imperfect CSI.}
\vspace{-0.6cm}
\section*{Appendix~A: Proof of Theorem~\ref{rank one relax}} \label{Appendix:A}
\vspace{-0.2cm}
Problem (P3.2) is a convex problem. Therefore, the optimal solution is characterized by the KKT conditions. Specifically, the Lagrangian function of Problem (P3.2) in terms of the active beamforming ${{{\mathbf{W}}_k}}$ can be written as
\vspace{-0.4cm}
\begin{align}\label{Lagrangian}
\begin{gathered}
  {\mathcal{L}} = \sum\limits_{j = 1}^K {\sum\limits_{k \ge j}^K {{\lambda _{kj}}\left( {{\rm {Tr}}\left( {{{\mathbf{W}}_k}{\mathbf{H}}_j^H{\mathbf{v}}{{\mathbf{v}}^H}{{\mathbf{H}}_j}} \right) - \frac{1}{{{S_{kj}}}}} \right)} }  \hfill \\
   + \sum\limits_{k = 1}^K {\sum\limits_{i < k}^K {\beta _{ki}^j\left( {{\rm {Tr}}\left( {{{\mathbf{W}}_i}{\mathbf{H}}_j^H{\mathbf{v}}{{\mathbf{v}}^H}{{\mathbf{H}}_j}} \right) + {\sigma ^2} - {I_{kj}}} \right)} }  \hfill \\
   + \sum\limits_{j = 1}^K {\sum\limits_{i = 1}^K {\sum\limits_{k \ge i}^K {\mu _{ki}^j\left( {{\rm {Tr}}\left( {{{\mathbf{W}}_k}{\mathbf{H}}_j^H{\mathbf{v}}{{\mathbf{v}}^H}{{\mathbf{H}}_j}} \right) - {\rm {Tr}}\left( {{{\mathbf{W}}_i}{\mathbf{H}}_j^H{\mathbf{v}}{{\mathbf{v}}^H}{{\mathbf{H}}_j}} \right)} \right)} } }  \hfill \\
   + \alpha \left( {{P_T} - \sum\limits_{k = 1}^K {{\rm {Tr}}\left( {{{\mathbf{W}}_k}} \right)} } \right) + \sum\limits_{k = 1}^K {{\rm {Tr}}\left( {{{\mathbf{W}}_k}{{\mathbf{Y}}_k}} \right)} + \mathbf{T} ,  \hfill \\
\end{gathered}
\end{align}
\vspace{-0.4cm}

\noindent where $\mathbf{T}$ denotes the terms which are independent of ${{\mathbf{W}}_k}$. ${{\lambda _{kj}}}$, ${\beta _{ki}^j}$, ${\mu _{ki}^j}$, $\alpha $ and ${{{\mathbf{Y}}_k}}$ are Lagrange multipliers associated with the corresponding constraints \eqref{S_kj P3.1}, \eqref{I_kj P3.1}, \eqref{fairness P3.1}, \eqref{total power W} and \eqref{SDP W}. The KKT conditions for the optimal ${\mathbf{W}}_k^*$ can be expressed as
\vspace{-0.3cm}
\begin{align}\label{KKT}
\begin{gathered}
  \lambda _{kj}^*,\beta _{ki}^{j*},\mu _{ki}^{j*},\alpha ^{*}  \ge 0,\;\;\;{\mathbf{Y}}_k^*{ \succeq }{\mathbf{0}} \hfill \\
  {\mathbf{Y}}_k^*{\mathbf{W}}_k^* = {\mathbf{0}},\;\;\;\;{\nabla _{{\mathbf{W}}_k^*}}{\mathcal{L}} = 0, \hfill \\
\end{gathered}
\end{align}
\vspace{-0.8cm}

\noindent where $\lambda _{kj}^*,\beta _{ki}^{j*},\mu _{ki}^{j*},\alpha ^{*}$ and ${\mathbf{Y}}_k^*$ denote the optimal Lagrange multipliers and ${\nabla _{{\mathbf{W}}_k^*}}{\mathcal{L}}$ represents the gradient of ${\mathcal{L}}$ with respect to ${\mathbf{W}}_k^*$. Then, the condition ${\nabla _{{\mathbf{W}}_k^*}}{\mathcal{L}} = 0$ can be expressed as
\vspace{-0.3cm}
\begin{align}\label{KKT=0}
{\alpha ^*}{\mathbf{I}} = {\mathbf{Y}}_k^* + {\mathbf{Zv}}{{\mathbf{v}}^H}{{\mathbf{Z}}^H},
\end{align}
\vspace{-1.0cm}

\noindent where ${\mathbf{Z}} $ is given by
\vspace{-0.3cm}
\begin{align}\label{Gamma}
{\mathbf{Z}} = \sum\limits_{j \le k}^{} {\lambda _{kj}^*} {\mathbf{H}}_j^H + \sum\limits_{j = 1}^K {\sum\limits_{i \le k}^{} {\mu _{ki}^{j*}{\mathbf{H}}_j^H} }.
\end{align}
\vspace{-0.8cm}

\indent Multiplying both sides of \textcolor{black}{\eqref{KKT=0}} by ${\mathbf{W}}_k^*$ and recalling that ${\mathbf{Y}}_k^*{\mathbf{W}}_k^* = {\mathbf{0}}$, we have ${\alpha ^*}{\mathbf{W}}_k^* = {\mathbf{Zv}}{{\mathbf{v}}^H}{{\mathbf{Z}}^H}{\mathbf{W}}_k^*$ and ${\alpha ^*}$ is always positive. With basic rank inequalities for matrices, we have
\vspace{-0.3cm}
\[
  {\rm {rank}}\left( {{\mathbf{W}}_k^*} \right) = {\rm {rank}}\left( {{\alpha ^*}{\mathbf{W}}_k^*} \right) = {\rm {rank}}\left( {{\mathbf{Zv}}{{\mathbf{v}}^H}{{\mathbf{Z}}^H}{\mathbf{W}}_k^*} \right) \le {\rm {rank}}\left( {{\mathbf{v}}{{\mathbf{v}}^H}} \right) = 1.
\]
\vspace{-1.0cm}

\noindent Therefore, ${\rm {rank}}\left( {{\mathbf{W}}_k^*} \right) \le 1$. The proof is completed.
\vspace{-0.6cm}
\section*{Appendix~B: SROCR approach framework} \label{Appendix:A}
\vspace{-0.2cm}
We present a brief review of the SROCR approach in a general framework. Instead of ignoring the rank-one constraint, the main idea of the SROCR approach is that the rank-one constraint is relaxed gradually to find a feasible rank-one solution. Consider the following problem
\vspace{-0.6cm}
\begin{subequations}
\begin{align}\label{P}
({{\rm{P}}}):&\mathop {\min }\limits_{{\mathbf{X}} \succeq 0} \;\;{g_0}\left( {\mathbf{X}} \right) \\
\label{P gx}{\rm{s.t.}}\;\;&{g_k}\left( {\mathbf{X}} \right){ \unlhd _k} 0,k = 1,2, \cdots ,K,\\
\label{rank 1 X}&{\rm {rank}}\left( {{{\mathbf{X}}}} \right) = 1,
\end{align}
\end{subequations}
\vspace{-1.2cm}

\noindent where ${g_k}:{{\mathbb{C}}^{N \times N}} \to {\mathbb{R}},k = 0,1,2, \cdots ,K$ are continuous and differentiable convex or affine functions with respect to a complex-valued positive semidefinite matrix variable ${{\mathbf{X}} \succeq 0}$ with the space of ${N \times N}$, and ${ \unlhd _k}$ denotes "$ \le $" or "$ = $". In order to handle the rank-one constraint \eqref{rank 1 X}, we have the following problem
\vspace{-0.6cm}
\begin{subequations}
\begin{align}\label{P w}
({{\rm{P}},\omega }):&\mathop {\min }\limits_{{\mathbf{X}} \succeq 0} \;\;{g_0}\left( {\mathbf{X}} \right) \\
\label{P gx}{\rm{s.t.}}\;\;&{g_k}\left( {\mathbf{X}} \right){ \unlhd _k} 0,k = 1,2, \cdots ,K,\\
\label{rank 1 w}&{\lambda _{\max }}\left( {\mathbf{X}} \right) \ge {\omega ^{\left( i \right)}}{\rm {Tr}}\left( {\mathbf{X}} \right),
\end{align}
\end{subequations}
\begin{algorithm}[t!]\label{method1}
\caption{SROCR algorithm for Problem (P)} 
 \hspace*{0.02in} 用来控制位置，同时利用 \\ 进行换行
\hspace*{0.02in}{Initialize convergence thresholds $\epsilon_1$, $\epsilon_2$, $i=0$.\\ Solve the relaxed problem $({\rm{P}},\omega )$ with ${\omega ^{\left( i \right)}} = 0$, the obtained solution is denoted by ${{{\mathbf{X}}^{\left( i \right)}}}$.\\ Initialize the step size ${\delta ^{\left( i \right)}} \in \left( {0,1 - {{{\lambda _{\max }}\left( {{{\mathbf{X}}^{\left( i \right)}}} \right)} \mathord{\left/
 {\vphantom {{{\lambda _{\max }}\left( {{{\mathbf{X}}^{\left( i \right)}}} \right)} {{\rm {Tr}}\left( {{{\mathbf{X}}^{\left( i \right)}}} \right)}}} \right.
 \kern-\nulldelimiterspace} {{\rm {Tr}}\left( {{{\mathbf{X}}^{\left( i \right)}}} \right)}}} \right]$.}\\
\vspace{-0.4cm}
\begin{algorithmic}[1]
\STATE {\bf repeat}
\STATE Solve the convex Problem $({\rm{P}},\omega )$ for given $\left\{ {{\omega ^{\left( i \right)}},{{\mathbf{X}}^{\left( i \right)}}} \right\}$.
\STATE {\bf if} Problem $({\rm{P}},\omega )$ is solvable {\bf then}
\STATE The optimal solution is denoted by ${{{\mathbf{X}}^{\left( {i + 1} \right)}}}$.
\STATE ${\delta ^{\left( {i + 1} \right)}} = {\delta ^{\left( i \right)}}$.
\STATE {\bf else}
\STATE ${\delta ^{\left( {i + 1} \right)}} = {{{\delta ^{\left( i \right)}}} \mathord{\left/
 {\vphantom {{{\delta ^{\left( i \right)}}} 2}} \right.
 \kern-\nulldelimiterspace} 2}$.
\STATE {\bf end}
\STATE ${\omega ^{\left( {i + 1} \right)}} = \min \left( {1,\frac{{{\lambda _{\max }}\left( {{{\mathbf{X}}^{\left( {i + 1} \right)}}} \right)}}{{{\rm {Tr}}\left( {{{\mathbf{X}}^{\left( {i + 1} \right)}}} \right)}} + {\delta ^{\left( {i + 1} \right)}}} \right)$.
\STATE $i=i+1$.
\STATE {\bf until} $\left| {1 - {\omega ^{\left( {i - 1} \right)}}} \right| \le \epsilon_1$ and $\left| {{g_0}\left( {{{\mathbf{X}}^{\left( i \right)}}} \right) - {g_0}\left( {{{\mathbf{X}}^{\left( {i - 1} \right)}}} \right)} \right| \le \epsilon_2$.
\end{algorithmic}
\end{algorithm}
\vspace{-1.2cm}

\noindent where ${\lambda _{\max }}\left( {\mathbf{X}} \right)$ denotes the largest eigenvalue of ${\mathbf{X}}$. In particular, if ${\omega ^{\left( i \right)}} = 1$, solving the above problem is able to find a rank-one solution to Problem (P). When ${\omega ^{\left( i \right)}} = 0$, the above problem is equivalent to ignoring the rank-one constraint as assumed in SDR approach. Motivated by this, we can increase ${\omega ^{\left( i \right)}}$ sequentially from 0 to 1 through iterations to gradually approach a rank-one solution. Specifically, ${\lambda _{\max }}\left( {\mathbf{X}} \right)$ can be expressed as
\vspace{-0.4cm}
\begin{align}\label{lamda max}
{\lambda _{\max }}\left( {\mathbf{X}} \right) = {{\mathbf{u}}_{\max }}{\left( {{{\mathbf{X}}^{\left( i \right)}}} \right)^H}{\mathbf{X}}{{\mathbf{u}}_{\max }}\left( {{{\mathbf{X}}^{\left( i \right)}}} \right),
\end{align}
\vspace{-1.2cm}

\noindent where ${{\mathbf{u}}_{\max }}\left( {{{\mathbf{X}}^{\left( i \right)}}} \right)$ is the eigenvector corresponding to the largest eigenvalue of ${{{\mathbf{X}}^{\left( i \right)}}}$ and ${{{\mathbf{X}}^{\left( i \right)}}}$ is the obtained feasible solution with parameter ${\omega ^{\left( i \right)}}$ from the previous iteration. The SROCR approach framework is summarized in \textbf{Algorithm 4}. Detailed discussion about the convergence can be found in~\cite{SROCR2017}, where it demonstrates that the sequence generated by the SROCR approach converges to a KKT stationary point of problem (P).

\bibliographystyle{IEEEtran}
\bibliography{mybib}

\end{document}